\documentclass[final,5p,times,twocolumn,authoryear]{elsarticle}
\usepackage{multirow}
\usepackage{graphicx}
\usepackage{amssymb}
\usepackage{amsthm}
\usepackage{lipsum}
\usepackage{xcolor}
\usepackage{comment}
\usepackage{mathtools}
\usepackage{gensymb}
\usepackage{floatrow}
\usepackage{dblfloatfix}
\usepackage{graphicx}
\usepackage{commath}
\usepackage{framed}
\usepackage{nomencl}
\usepackage{siunitx}
\makenomenclature

\usepackage[label font=normal]{subfig}
\usepackage{caption}
\usepackage[ruled,vlined]{algorithm2e}
\SetKwProg{Init}{Initialization:}{}{}
\theoremstyle{definition}

\usepackage{xcolor}
\usepackage[colorlinks=true]{hyperref}
	\hypersetup{
		linkcolor=blue,
		citecolor=red,
		urlcolor=green
	}

\usepackage{pifont}
\usepackage{diagbox}
\usepackage{enumerate}
\usepackage{mathrsfs}
\usepackage{color}
\usepackage{enumerate,graphicx,epstopdf}
\usepackage{amssymb,bm}
\usepackage{mathtools}
\usepackage{array}
\usepackage{algpseudocode,algorithmicx}
\usepackage{booktabs}
\usepackage{graphicx}
\usepackage{caption}
\usepackage{amsmath}
\usepackage{siunitx}
\usepackage{algpseudocode}
\usepackage{lipsum}

\begin{document}
\begin{frontmatter}
\title{\LARGE Plug and Play Distributed Control of Clustered Energy Hub Networks}

\address[label1]{Automatic Control Laboratory, Swiss Federal Institute of Technology (ETH), Z\"{u}rich, Switzerland}
\address[label2]{Urban Energy Systems Laboratory, Swiss Federal Laboratories for Materials Science and Technology (Empa), D\"{u}bendorf, Switzerland}
\address[label3]{Chair of Mathematical Systems Theory, Department of Mathematics,  University of Stuttgart, Stuttgart, Germany}

\author[label1,label2]{Varsha N. Behrunani\, \corref{cor1}}
\author[label1,label2]{Cara Koepele\, \corref{cor1}}
\author[label1,label3]{Jared Miller\, }
\author[label1]{Ahmed Aboudonia\, }
\author[label2]{Philipp Heer\, }
\author[label1]{Roy S. Smith\, }
\author[label1]{John Lygeros}

\cortext[cor1]{These authors contributed equally to the work.}

%%%%%%%%%%%%%%%%%%%%%%%%%%%%%%%%%%%%%%%%%%
% Abstract:
\begin{abstract}

The transition to renewable energy is driving the rise of distributed multi-energy systems, in which individual energy hubs and prosumers (e.g., homes, industrial campuses) generate, store, and trade energy. Economic Model Predictive Control (MPC) schemes are widely used to optimize operation of energy hubs by efficiently dispatching resources and minimizing costs while ensuring operational constraints are met. Peer-to-peer (P2P) energy trading among hubs enhances network efficiency and reduces costs but also increases computational and privacy challenges, especially as the network scales. Additionally, current distributed control techniques require global recomputation whenever the network topology changes, limiting scalability. To address these challenges, we propose a clustering-based P2P trading framework that enables plug-and-play operation, allowing energy hubs to seamlessly join or leave without requiring network-wide controller updates. The impact is restricted to the hubs within the affected cluster. The energy trading problem is formulated as a bi-level bargaining game, where inter-cluster trading commitments are determined at the cluster level, while energy dispatch and cost-sharing among hubs within a cluster are refined at the hub level. Both levels are solved in a distributed manner using ADMM, ensuring computational feasibility and privacy preservation. Moreover, we develop plug-and-play procedures to handle dynamic topology changes at both the hub and cluster levels, minimizing disruptions across the network. Simulation results demonstrate that the proposed bi-level framework reduces operational costs, and enables scalable energy management under plug-and-play operation.

\end{abstract}
%%%%%%%%%%%%%%%%%%%%%%%%%%%%%%%%%%%%%%%%%%
% Abstract:
\end{frontmatter}
\section{Introduction}
\label{chp:1_Introduction}

% \begin{enumerate}
%     \item Context and Motivation (changing landscape, sustainability)
%     \item Energy Hubs intro and goals surrounding their operation and markets
%     \item previous work and challenges, particularly in the landscape of large networks (i.e. needing a central coordinator), introduce plug-and-play.
%     \item Research proposal and why
%     \item report structure
% \end{enumerate}

% Summary: An increase in DERs means that there are many new coordination possibilities to consider (sets the tone for large networks)
With the influx of distributed energy resources (DERs) such as rooftop solar PV, electric batteries, and thermal storage, individual consumers can now generate and store energy in addition to purchase power from the grid. The role of the consumer is thus shifting to a `prosumer,' where individual users can both produce and consume energy. The presence of prosumers is one factor marking the shift of the energy grid from a centralized, top-down architecture \citep{kristov2016tale} to a distributed architecture with behind-the-meter generation. Planning and control for distributed energy generation introduces new coordination challenges between increasingly interconnected energy sources and loads that were not previously present in the centralized generation model. 

% Summary: Explanation of an energy hub, why it is useful, and the baseline algorithms for single hubs.
The collection of loads, storage systems, and generation technologies present at an individual site can be described as an energy hub \citep{geidl2006ehubs}. The energy hub manages its devices to fulfill the time-varying demands across multiple energy carriers. This joint hub-focused management allows for more flexible and efficient operation compared to operating each device independently. Advanced algorithms such as optimal power flow \citep{almassalkhi:2011z}, \citep{geng:2020r} and Model Predictive Control (MPC) \citep{arnold:2009h}, \citep{chandan:2014j} have been explored as control strategies for single energy hubs. These algorithms can consider multiple factors such as the cost of operation, carbon intensity, obedience of thermal comfort constraints, and fidelity to other structural constraints. 

% Summary: If trade is allowed, it needs to be fair to motivate agents to stay in the market.
%If infrastructural connections exist between different energy hubs, the hubs may trade electricity, gas, and heat with each other rather than buying or generating these resources on their own. 
Peer-to-peer (P2P) energy trading between different energy hubs can further reduce the individual or aggregate cost of operation for energy hubs compared to the decentralized setting of no trading. % \cmr{Something about social welfare goes here}. 
% % Advanced control techniques (MPC in particular) for energy hub operation
% Model predictive control (MPC) is an advanced control technique which at each time instant solves a constrained optimal control problem over a prediction horizon and applies the first control action in the optimal control sequence. MPC is found to be a competitive contender for energy hub control design. This is attributed to its ability to handle the optimization requirements in a systematic manner and to explicitly incorporate constraints — whether hard or soft, applied to states or inputs. By solving the constrained optimal control problem at every timestep, MPC allows the energy hub to adapt to disturbances such as changing demands and weather conditions. These are often provided as forecasts, and MPC can modify the system's control strategy to the updated forecast. %This is also because MPC is characterized by its capability to make use of the predicted system behavior and disturbance forecast.
% Summary: A popular optimization scheme is MPC, and naturally this can be solved in a centralized manner but it has drawbacks.
Centralized MPC can be used to optimize the operation and trading of an entire energy hub network. Several centralized methods have been proposed in the literature, formulating the energy hub network problem as an optimization task~\citep{Smith:2022} to achieve specific objectives such as cost reduction, environmental benefits, and demand response~\cite{Scala:2014,Yang:2016, Maroufmashat:2015}. 
%MPC is usually designed and implemented in a centralized fashion.The operation of energy hub planning and trading can be performed through centralized or distributed control paradigms (solving associated optimization problems). Although centralized MPC approaches tend to maximize the social welfare of energy hubs involved in trading, such approaches suffer from poor scalability as the network size increases or changes,...
Though this strategy can minimize overall network costs, it suffers from poor scalability as the network size increases or changes, inhibiting the use of centralized algorithms in real-time operation. Advanced algorithms must be used to address these computational challenges~\cite{Aghtaie:2014}. Centralized approaches also compromise privacy, as they require detailed per-agent system and state information for policy computation. Furthermore, centralized control systems are more susceptible to failure, as a single control failure can lead to the collapse of the entire system. Finally, these control systems are not well-suited for networks with varying topologies, as even a small change in the network configuration necessitates the redesign of the entire control system.

% Summary: Why distributed, and the downsides of a central coordinator.
To tackle the aforementioned problems, multiple non-centralized MPC schemes have been developed over the years \citep{scattolini2009architectures,negenborn2014distributed}. These schemes are based on the idea of decomposing the system into several smaller subsystems and designing a local controller for each of these subsystems. Contrary to decentralized MPC,  distributed MPC approaches allow the local controller of each subsystem to communicate with the local controllers of some or all other subsystems in the network. These non-centralized schemes are usually characterized by their scalability, privacy and robustness against failure. %On the other hand, a centralized coordinator still exists and communicates with the local controllers of the different subsystems in hierarchical MPC approaches. 
% Summary: Distributed energy hub network algorithms exist already, but may not consider fairness and topology changes.
In the context of energy hubs, distributed MPC can be used to split the energy hub optimization into per-hub optimization problems. The agents locally solve their own problem, exchange state information or dual variables with other agents, and repeat this process until convergence of a control policy is achieved. Specific optimization methods for distributed energy hub MPC include ADMM \citep{behrunani2023distributed, WANG2020CoordeHubs}, proximal atomic coordination with Nesterov acceleration \citep{ferro2022PAC} and dual-regularized fast dual forward-backward splitting \citep{pelzmann2021buildings}. However, these methods do not guarantee that all agents will individually profit from this arrangement, which could motivate some hubs to leave the network. 
The trading scheme must satisfy properties of \textit{fairness}, otherwise agents may elect to leave the market and stop trading altogether. The trading mechanism and costs must be designed to ensure that hubs are encouraged to participate, such as by allocating cost savings in proportion with the agent's contribution to the network. To achieve this, energy hub trading in a network can be formulated as a game where each hub tries to maximize its own benefit. Various general welfare definitions for incorporating fairness into optimization problems are presented in \citep{chen2023Fairness}. The impact of P2P prices with respect to agent preferences and marginal prices (dual variables) is presented in \citep{lcadre2020p2p}. In \citep{behrunani2023fairness}, a distributed bi-level game determines optimal setpoints and trading prices that yield equal proportional cost reduction for all network participants. The authors of \citep{WEI2017Stackelberg} define resource prices as the Stackelberg Equilibrium of a multi-leader multi-follower Stackelberg Game between distributed energy providers and energy users. The solution point is found using a distributed Best Response algorithm based on the utility functions of the agents. A distributed bargaining game with a central coordinator determines net electricity trades and compensation in \citep{Fan2018Bargaining} which causes equal absolute financial benefit for each energy hub.

Furthermore, distributed control schemes require agents to explicitly solve for energy trades with every other agent in the network, meaning that topology changes would require all agents in the network to reformulate their optimal control problem. %Summary: There are benefits to combining individual energy hubs together to protect against impacts of topology changes 
Energy hubs can be grouped into clusters to limit the impact of topology changes and the size of optimization problems for distributed control strategies. This is a similar concept to an aggregator \citep{burger2017aggregators}, which combines the resources from a cluster of energy hubs in order to participate in high-level market trading \citep{di2018optimal}. This allows individual hubs within an aggregator to trade among themselves without considering the high-level structure. The authors in \citep{chen2023multicluster} determine resource prices, and optimal dispatch in a large-scale energy system with an aggregative game between multiple subnets (clusters). The prices are dependent on the aggregate actions of all network participants. Their method shows promise in scalability and flexibility for different topologies, but the network model does not consider grid or energy hub constraints. The work of \citep{chen2023clusteredstackelberg}, formulates a Stackelberg game in which multiple cluster coordinators set prices for prosumer followers in their cluster.

% With a clustered architecture, the impact of an entering or exiting participant is limited to the cluster. The authors in \citep{chen2023multicluster} determine resource prices, generation, and consumption decisions in a large-scale energy system with an aggregative game between multiple subnets. The prices are dependent on the aggregate actions of all network participants. Their method shows promise in scalability and flexibility for different topologies, but the network model does not consider grid or energy hub constraints. In \citep{chen2023clusteredstackelberg}, a Stackelberg game is formulated such that multiple region coordinators set prices for prosumer followers in their region.

% Summary: PnP procedures are useful to handle likely changes in the ehub network and have been employed in other fields.
The distributed MPC setting requires additional guarantees in order to tolerate time-evolving topologies, such as removals, migrations, or joins of energy hubs in the network. The Plug-and-Play (PnP) framework \citep{stoustrup2009plug} offers a general guideline for updating control policies changes in the network structure. After detecting a network change, a proper PnP procedure decides whether to accept the change and uses local information to update its controllers. Decentralized PnP approaches were developed in \citep{riverso2013PnPDeMPC} where all couplings were considered as disturbances. Additionally, two-phase distributed PnP approaches were also developed in \citep{zeilinger2013plug}. %First, a redeisgn phase was implemented to ensure the closed-loop stability of the new network after the PnP operation by redesigning the local controllers of the plugged-in and plugged-out subsystems as well as their agents. Second, a transition phase was implemented to ensure the recursive feasibility by driving the system to a feasible steady state at which the PnP operation takes place. Since then, 
Distributed PnP approaches have been used in many applications including electric vehicle charging \citep{bansal2014plug}, smart grids \citep{le2017plug} and vehicle platoons \citep{hu2018plug}. The authors in \citep{aboudonia2022refmpcpnp} also proposed distributed PnP procedures based on passivity theory to accelerate PnP operations and increase flexibility by accepting more requests. Clustering offers a natural formulation for realizing PnP practices in energy hubs: when an energy hub joins or leaves a cluster, energy hubs outside of that cluster should not need to recompute their policies.

Here, we introduce a cluster-based scheme for fair energy hub trading that is compatible with PnP operation. The proposed control architecture involves two levels: the upper cluster level (inter-cluster) and the lower hub level (intra-cluster). Within each cluster at the hub level, each energy hub computes their respective consumption and generation with respect to trading among other hubs in the same cluster. At the cluster level, each cluster computes import and export trades between other clusters. The cluster level recomputes trades less frequently than the hub level to minimize computational burden. The trades between clusters are fixed such that the hub level of each cluster can operate independently between cluster level computations. The costs are also rebalanced in each cluster to maintain fairness, thus ensuring that remaining in the market is favorable for every energy hub in each cluster. The contributions of this work are as follows:

% \cmr{Original introduction of Cara's report. Going to be rewriting some of this.}

% In this project, we enable the plug-and-play of fully modeled energy hubs into a control scheme by aggregating agents into clusters and isolating the impact of an entering or exiting hub to their cluster. Entering and exiting clusters only impact the communication schemes of one hub per cluster. A distributed bi-level Nash Bargaining Game (NBG) algorithm is formulated at the cluster level to determine aggregated trades and fair compensation. These sub-networks of hubs determine their trades hourly with MPC while fulfilling their cluster trade commitment. Finally, a cost-balancing step in each cluster ensures that the cluster compensation is divided fairly among its energy hubs. 

\begin{enumerate}
    \item We formulate a Nash bargaining game between energy hub clusters without a central network coordinator to determine trades and cost bids between clusters. 
    \item We develop a novel distributed receding-horizon control strategy for a network of clustered energy hubs to solve the bargaining game between clusters and suitably divides the cluster trades between the hubs in each cluster.
    \item We define PnP procedures for the proposed algorithm to handle hubs and clusters joining and leaving the network.
    \item We formulate and analyze the cost-fairness methodology that distributes the cluster benefit between hubs in each cluster under receding horizon control.
\end{enumerate}

Finally, we demonstrate near-optimal behavior of the proposed algorithm under static and changing network topologies and varying conditions with numerical simulations on a multi-hub network, using realistic models of energy hubs and demand data. This paper is organized as follows: Section \ref{sec:ehub} explains the energy hub network model. Section \ref{sec:clustering} accumulates the energy hubs into clusters, and formulates inter-cluster and intra-cluster optimization problems. Section \ref{sec:opt} presents our proposed optimization framework for clustered energy hubs. Section \ref{sec:pnp} explains how the clustering optimization problems can accommodate plug-and-play operations of hubs/clusters joining or leaving the network. Section \ref{sec:experiments} demonstrates our clustering-based energy hub trading-enabled MPC on simulated examples. Section \ref{sec:conclusion} concludes the paper.

% The main contribution is the development of the control scheme along with PnP procedures for hubs and clusters entering or exiting the network. The performance of the developed algorithm is demonstrated in simulation under static and changing network topologies.

% Notation?
% \clearpage
%\section{Preliminaries} \label{sec:preliminaries}
\subsection{Notation}
The set of real numbers and integers are denoted by $\mathbb{R}$ and $\mathbb{Z}$, respectively. $\mathbb{Z}_{[a,b]}$ denotes the set of numbers $\{a,a+1,...,b\}$. The set of all integers $\geq a$ is denoted $\mathbb{Z}_{\geq a}$. $\mathbb{R}^n$ denotes the $n$-dimensional real vector space. 

\section{Modelling and optimization of energy hub network}
\label{sec:ehub}
\begin{figure}[t]
\centering
\includegraphics[width=89mm]{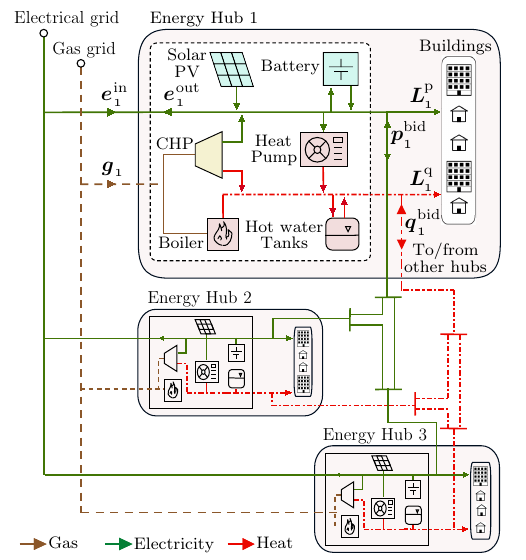}
  \caption{A network of three interconnected energy hubs. Each hub can import energy from the electricity and gas grids, and can feed-in electricity to the electricity grid. Additionally, each hub can also trade electrical and thermal energy with the other hubs.}
  \label{fig:elect_grid_hub}
 \end{figure} 

We consider a network of $H$ interconnected energy hubs indexed by $i \in \mathcal{H} := \{1,\dots,H\}$. Each hub comprises different generation, storage and conversion devices and is connected to main electricity and gas grids. It utilizes energy from the grids to fulfill the electricity and thermal demands of the buildings connected to it. Excess electrical energy produced in the hub can be fed back into the grid. Hubs in the network can also trade electrical energy using the existing electricity grid, and trade heat via a local thermal grid. The electricity and heat demands at each time are considered to be uncontrolled, and act as a disturbance for the hub controller. In this study, we assume that a perfect forecast is available for this demand. Hub devices include renewable generation sources such as solar photovoltaic and solar thermal collectors, conventional generation units such as Combined Heat and Power (CHP), and conversion devices such as gas boilers and heat pumps, that generate electricity and/or heat thermal energy. Hubs may also include different storage devices such as batteries to store electrical energy and water tanks to store thermal energy Fig. \ref{fig:elect_grid_hub} shows an example network of three interconnected energy hubs and the details of an example hub in the network.

\subsection{Energy hub modeling}
In this section, we model the energy hub network and formulate the corresponding optimal control problem. Consider a horizon $\mathcal{T} = \{0,\dots,T-1\}$ of length $T$. Let $\mathcal{N}_i$ denote the set of devices constituting the energy hub $i$. The dynamics of each energy hub device $n \in \mathcal{N}_i$ are modeled as the following discrete-time linear state-space system that describes the evolution of its internal energy state $x_{t,n}$:
\begin{equation}
\label{eq:hub_dynamics}
\left.\begin{array}{c}
\begin{aligned}
x_{t+1,n}&=A_{n} x_{t,n} + B_{n} u_{t,n} + D_{n}  d_{t,n}, \\
x_{t,n} &\in \mathcal{X}_{n}, \ 
u_{t,n} \in \mathcal{U}_{n},
\end{aligned}
\end{array}\right\} \forall t \in \mathcal{T},
\end{equation}
where $d_{t,n}$ are the exogenous disturbances (e.g. solar radiation) acting on the system and $u_{t,n}$ is the vector of control inputs for device $n$ at time $t$. $A_{n},B_{n}$ and $D_{n}$ are the state, input and disturbance matrices, and $\mathcal{X}_{n}$ and $\mathcal{U}_{n}$ are the state and input constraint set, respectively. Here, we assume that a perfect forecast for the temperature and the solar radiation are available. The vector $u_{t,n}$ is defined as 
\begin{equation}
\nonumber
\begin{aligned}
\begin{array}{c c c }
u_{t,n} = 
\left[ \begin{array}{c c c c c}
u_{t,n}^{\mathrm{g,in}},& 
u_{t,n}^{\mathrm{p,in}},&
u_{t,n}^{\mathrm{q,in}},&
u_{t,n}^{\mathrm{p,out}},&
u_{t,n}^{\mathrm{q,out}}\end{array} \right]^{\text{T}},
\end{array}
\end{aligned}
\end{equation}
where $u_{t,n}^{\mathrm{g,in}}$, $u_{t,n}^{\mathrm{p,in}}$ and $u_{t,n}^{\mathrm{q,in}}$ are the gas, electricity and heating input to the device at time $t$, respectively, and $u_{t,n}^{\mathrm{p, out}}$ and $u_{t,n}^{\mathrm{q,out}}$ are the electricity and heating outputs from the device at time $t$, respectively. Given a time horizon $\mathcal{T} \subset \mathbb{Z}$, for any variable $\boldsymbol{v}_{t\mathrm{,a}}$ at time $t$, the vector variable $\boldsymbol{v}_{\mathrm{a}}$ collects all the variables over the complete horizon $\mathcal{T}$ such that $\boldsymbol{v}_{\mathrm{a}}=\{\boldsymbol{v}_{t\mathrm{,a}}\}_{t \in \mathcal{T}}$.

The energy hub internal network is characterized by the electricity, heating and gas energy balance constraints. The electricity energy balance constraint for hub $i$ is given by
\begin{subequations}
\label{eq:hub_balance}
\begin{equation}
\label{eq:hub_elecbalance}
\begin{aligned}
L^{\mathrm{p}}_{i} &= e^{\mathrm{out}}_{i} - e^{\mathrm{in}}_{i} + \sum_{n \in \mathcal{N}_{i}} \left(u_{n}^{\mathrm{p,out}} - u_{n}^{\mathrm{p,in}}\right) + \eta_p p_{i}^{\mathrm{bid,out}} - p_{i}^{\mathrm{bid,in}}
\end{aligned}
\end{equation}
where $L^{\mathrm{p}}_{i}$ is the total electricity demand of the consumers supplied by the energy hub $i$, and $e^{\mathrm{out}}_{i} \geq 0 $ and $e^{\mathrm{in}}_{i}\geq 0$ are the electricity purchased from and sold to the electricity grid, respectively. Here, $p_{i}^{\mathrm{bid,out}}$ and $p_{i}^{\mathrm{bid,in}}$ is the net energy imported from, and exported to the other hubs, respectively, and the losses through electrical lines are accounted for in the energy flowing into the hub using a blanket proportional efficiency factor, $\eta_{\mathrm{p}}$. The energy balance also includes the electrical energy input and output from the hub devices. Similarly, the thermal energy balance constraint for hub $i$ is given by
\begin{equation}
\label{eq:hub_heatbalance}
\begin{aligned}
L^{\mathrm{q}}_{i} &= \sum_{n \in \mathcal{N}_{i}} \left(u_{n}^{\mathrm{q,out}} - u_{n}^{\mathrm{q,in}}\right)  + \eta_q q_{i}^{\mathrm{bid,out}} - q_{i}^{\mathrm{bid,in}},
\end{aligned}
\end{equation}
where $L^{\mathrm{q}}_{i}$ is the total thermal demand served by the energy hub $i$, and $q_{i}^{\mathrm{bid,out}}$ and $q_{i}^{\mathrm{bid,in}}$ are the net thermal energy flow into and out of hub $i$ from/to the other hubs, respectively. The thermal losses are incorporated using efficiency factor, $\eta_{\mathrm{q}}$. We assume there is no global thermal grid and no thermal losses within each hub. The demand is fulfilled by the thermal input and output from the devices in the hub as well as through thermal energy traded with other energy hubs using a local heat distribution network. In the absence of P2P energy trades, the devices in each hub are sized such that the thermal demand can be completely met locally at all times by conversion or storage within the hub. A simplified model of the thermal dynamics within the hub is considered here. A more detailed model that considers temperature constraints, hydraulics, pipe dynamics, district heating, etc. is not considered here and left as a topic of future work. Finally, the net gas demand of the hub, $g_{i}$, is given by
\begin{equation}
\label{eq:hub_gasbalance}
\begin{aligned}
g_{i} = \sum_{n \in \mathcal{N}_{i}} u_{n}^{\mathrm{g,in}}.
\end{aligned}
\end{equation}
\end{subequations}

Let $p_{i}^{\mathrm{bid}}$ and $q_{i}^{\mathrm{bid}}$ be the net electrical and thermal energy that hub $i$ trades with the other hubs, respectively, which are defined as 
\begin{subequations}
\label{eq:hub_transfer}
\begin{equation}
\begin{aligned}
p_{i}^{\mathrm{bid}} &= p_{i}^{\mathrm{bid,out}} - p_{i}^{\mathrm{bid,in}},\\
q_{i}^{\mathrm{bid}} &= q_{i}^{\mathrm{bid,out}} - q_{i}^{\mathrm{bid,in}}.
\end{aligned}
\end{equation}
These variables are introduced for the energy balance over the network. Furthermore, additional constraints are imposed to limit the energy traded by the hubs,
\begin{equation}
\begin{aligned}
0&\leq p_{i}^{\mathrm{bid,in}}, \ p_{i}^{\mathrm{bid,out}}\leq \bar{p}^{\mathrm{bid}}, \\
0&\leq q_{i}^{\mathrm{bid,in}}, \ q_{i}^{\mathrm{bid,out}}\leq \bar{q}^{\mathrm{bid}}, 
\end{aligned}
\end{equation}
\end{subequations}
where $\bar{p}^{\mathrm{bid}}$ and $\bar{q}^{\mathrm{bid}}$ is the upper limit of the electrical and thermal energy that can be traded.
Let $p_{i}$ collect all the operational set points for the energy hub $i$, and $\mathscr{P}_{i}$ be the full constraint set given by: 
\begin{equation}
\nonumber
p_{i} = \big\{ \left\{u_{n}, x_{n}\right\}_{\forall n \in \mathcal{N}_i},\ e^{\mathrm{in/out}}_{i}, \ g_{i}, \ p_{i}^{\mathrm{bid}} , \  p_{i}^{\mathrm{bid,in/out}} , \ 
q_{i}^{\mathrm{bid}}, \ q_{i}^{\mathrm{bid,in/out}} \big\} ,
\end{equation}
\begin{equation}
\nonumber
    \mathscr{P}_{i}:= \left \{ p_{i}~|~ \text{(\ref{eq:hub_dynamics}, \ref{eq:hub_balance}, \ref{eq:hub_transfer}) hold}\right \}.
\end{equation}
\subsection{Optimization of the energy hub network}
In this study, the energy hub optimization aims to minimize the total operational costs while ensuring constraint satisfaction over the complete horizon $\mathcal{T}$. In the absence of any trading, this is the sum of the cost of the energy exchanged with the electricity and gas grids. The resulting decentralized economic dispatch optimization problem is compactly written as:
\begin{align}
    \min_{p_i} & \ 
    \underbrace{{{c^{\mathrm{out}}_{\mathrm{e}}}^T e^{\mathrm{out}}_{i} 
    - c^{\mathrm{in}}_{\mathrm{e}}}^T e^{\mathrm{in}}_{i} 
    + {c^{\mathrm{out}}_{\mathrm{g}}}^T g_{i}}_{J_{\mathrm{dec},i}} \notag \\
    \text{s.t.} & \quad p_i \in \mathscr{P}_i,\label{eq:dec_energyhub_optimization}\\
    \notag & \quad \bar{p}_{i}^{\mathrm{bid}} , \ \bar{q}_{i}^{\mathrm{bid}} = 0.
\end{align}
where $c^{\mathrm{out}}_{\mathrm{e}}$ and $c^{\mathrm{out}}_{\mathrm{g}}$ are the per unit price vectors for purchasing electricity and natural gas from the public grids, and $c^{\mathrm{in}}_{\mathrm{e}}$ is the feed-in tariff for the electricity grid. We assume dynamic prices for electricity and gas that vary at each time $t$, are independent of the aggregate demands of the network, and is known by the hub controller for the complete horizon $\mathcal{T}$. Considering uncertainty in pricing and demand-dependent prices is left for future work. Furthermore, while electricity prices may differ across hubs to reflect locational marginal pricing or specific contracts with grid operators in practice, we assume uniform prices for all hubs in the network here for simplicity. If the energy hub trades energy with other hubs, the cost of trading for hub $i$ is given by:
\begin{equation}
   \notag {c^{\mathrm{tr}}_{\mathrm{e}}}^T\lvert p_{i}^{\mathrm{bid}}\rvert + c_i^{\mathrm{bid}},
\end{equation}
where $c^{\mathrm{tr}}_{\mathrm{e}}$ is a trading tariff imposed by the grid operator for the use of the network for P2P trading, and $c_i^{\mathrm{bid}}$ is a scalar value for the net cost incurred by the hub $i$ for the total energy traded with the other hubs over the complete horizon $\mathcal{T}$. This value is positive if the hub pays for the energy trades and negative if the hub receives payment from the other hubs. The resulting cost objective for the complete network is the sum of the decentralized costs and trading costs for each of the hubs in the network. The resulting centralized optimization problem of the complete network is written as:
\begin{align}
    \min_{\{p_i,c_i^{\mathrm{bid}}\}_{\forall i \in \mathcal{H}}} & \ \sum_{i\in\mathcal{H}} \bigg(
    \underbrace{{{c^{\mathrm{out}}_{\mathrm{e}}}^T e^{\mathrm{out}}_{i}  
    - c^{\mathrm{in}}_{\mathrm{e}}}^T  e^{\mathrm{in}}_{i}
    + {c^{\mathrm{out}}_{\mathrm{g}}}^T g_{i}
    + {c^{\mathrm{tr}}_{\mathrm{e}}}^T\lvert p_{i}^{\mathrm{bid}}\rvert}_{J_{\mathrm{grid},i}}\bigg) \notag \\
    \text{s.t.} & \quad p_i \in \mathscr{P}_i, \ \  \forall i\in\mathcal{H} \label{eq:energyhub_optimization}\\
       \notag   &\quad\sum_{i\in\mathcal{H}}{p}_{i}^{\mathrm{bid}}=0,\\
       \notag & \quad\sum_{i\in\mathcal{H}}{q}_{i}^{\mathrm{bid}}=0, \\
      \notag &\quad \sum_{i\in\mathcal{H}}{c}_{i}^{\mathrm{bid}} =0. 
\end{align}

The resulting cost function no longer contains the ${c}_{i}^{\mathrm{bid}}$. This is because the sum of trade costs for the complete network is 0, that is,  $\sum_{i\in\mathcal{H}}{c}_{i}^{\mathrm{bid}} =0$ as the trading costs are exchanged with the hubs in the network. Furthermore, the sum of all the electrical and thermal energy traded within the network is also 0 since the trade agreements do not consider any network energy losses. These losses are instead incorporated in the energy balance equations of the hubs by using the efficiency factors. This also ensures no cyclic trades occur. %We consider that the thermal trading can only occur within the immediate geographical neighborhood. 

\section{Distributed optimization of a clustered energy hub network} 
\label{sec:clustering}

While the centralized optimization problem computes the global optimum for the entire system, it comes with scalability and privacy concerns. Centralized controllers are also more susceptible to topology changes and have to be reconfigured each time a hub enters or leave the network, which can be tedious for large networks. Finally, the centralized optimization of the form \eqref{eq:energyhub_optimization} maximizes the social welfare of the network and does not include any strategy to compute the hub trading costs. 

\begin{figure}[!t]
\centering
\includegraphics[width=89mm]{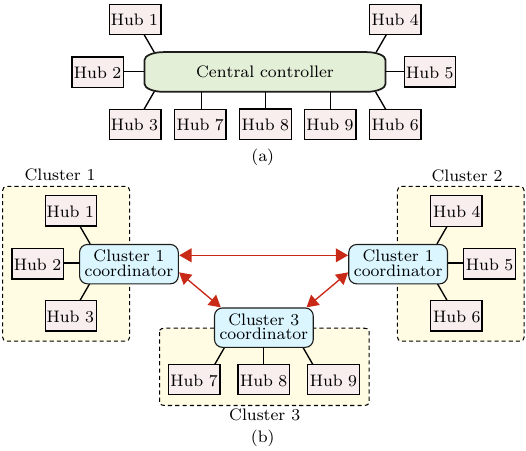}
  \caption{An energy hub network with the (a) centralized framework and (b) proposed framework where the hubs are divided into clusters each with a virtual cluster coordinator. The red lines depict the inter-cluster communication between cluster coordinators and the blue lines indicate the intra-cluster communication between the cluster coordinator and the hubs in the cluster.}
  \label{fig:network_cluster}
 \end{figure}

To mitigate these issues, we group the energy hubs in the network into smaller groups referred to here as "clusters" that can operate independently from each other. Fig. \ref{fig:network_cluster} depicts the communication and control strategy for the centralized network framework compared to the clustered network framework. We propose a hierarchical bargaining game algorithm to optimize the operation, trades, and costs in the complete network. In the outer loop, the clusters solve a bargaining game to optimize the energy trades and trading costs between them. In the inner loop, the hubs within each cluster compute the optimal hub setpoints as well as the hub trades in accordance to the energy trades computed in the outer loop. This strategy allows the computational load to be distributed. To achieve this distributed computation, each cluster contains a virtual intermediary referred to here as a "cluster coordinator" to compute the energy traded between clusters in the outer loop game as well as to coordinate the trades between the hubs within the cluster. The trades between the clusters in the upper level are computed less frequently and remain fixed while the hubs within each cluster can optimize more frequently in a receding horizon fashion. 
%In general, energy hub device models and constraints may be non-convex, and the resulting optimization problem can be large, making real-time solutions computationally prohibitive. The centralized controller needs to collect detailed information about the demand characteristics, model, and current state of all converters in each hub, information that the hub operators may not want to divulge. Centralized controllers are also more susceptible to topology changes and have to be reconfigured each time a hub enters or leave the network, which can be tedious for large networks. Finally, the centralized optimization of the form \eqref{eq:energyhub_optimization} maximizes the social welfare of the network and does not include any strategy to compute the hub trading costs. Hence, the optimization does not account for individual hub objectives which may result in an unfair outcome where not all hubs benefit from such a cooperative operation. 

Let the hubs be grouped into $M$ clusters indexed by $m \in \mathcal{M} := \{1,\dots,M\}$, and let the set of hubs in cluster $m$ be $\mathcal{H}_m$. The cluster coordinator of each cluster $m$ needs to compute both the cluster trade, $P^{\mathrm{bid}}_m$, and the corresponding cluster cost bid, $C^{\mathrm{bid}}_m$, for trading with the other clusters in the network. In order to compute the optimal solution with a fair distribution of the benefits of trading, we use the Nash Bargaining Game which optimizes the entire community performance whilst guaranteeing that the surplus profits of cooperation are equally distributed among the participants. The solution of a Nash Bargaining Game maximizes the welfare function (WF): 
\begin{equation}
    \notag \text{WF} = \max \prod_{m\in\mathcal{M}}(J^{\mathrm{dec}}_m - J_m)  = \max\prod_{m\in\mathcal{M}}(\underbrace{J^{\mathrm{dec}}_m - (J^{\mathrm{grid}}_m }_{\Delta J_m}+ C^{\mathrm{bid}}_m )).
\end{equation}
The optimizer of this WF also minimizes the convex objective function,
\begin{equation}
   \notag  J^{\mathrm{nbg}} = \min \sum_{m\in\mathcal{M}}-\ln{(\Delta J_m - C^{\mathrm{bid}}_m)},
\end{equation}
where $J^{\mathrm{dec}}_m$ is the net decentralized cost of the cluster $m$ in the absence of trading when all hubs in the cluster operate independently based on \eqref{eq:dec_energyhub_optimization}, $J_m$ is the cost of cluster $m$ with P2P trading which is the sum of the grid cost of the cluster from \eqref{eq:energyhub_optimization}, $J^{\mathrm{grid}}_m$, and the cost bid of the cluster, $C^{\mathrm{bid}}_m$, and $\Delta J_m$ is the difference between the decentralized and grid costs of the cluster in the absence and presence of P2P trading. For each cluster $m$, these costs are the sum of the costs of all the individual hubs in the cluster, defined as:
\begin{align}
 \notag   J^{\mathrm{dec}}_m &=  \sum_{i\in\mathcal{H}_m} J^{\mathrm{dec}}_i, \\  \notag  J^{\mathrm{grid}}_m &=  \sum_{i\in\mathcal{H}_m} J^{\mathrm{grid}}_i, \\  \notag  \Delta J_m &= J^{\mathrm{dec}}_m -  J^{\mathrm{grid}}_m = \sum_{i\in\mathcal{H}_m} \underbrace{\Delta J^{\mathrm{dec}}_i - J^{\mathrm{grid}}_i}_{J_i},
    \end{align}
where $\Delta J_i$ is the difference between decentralized cost and grid cost of the hub $i$ without and with trading. The standard bargaining game splits the benefits equally among all the agents regardless of the size and market contribution of the participating agents. In this study, we use a modified version of the bargaining game that weighs the objective of each cluster by a factor of $\alpha_m$ to distribute benefit based on cluster size. Cluster size can be defined by factors such as physical area, annual energy demand, or the number of hubs. Here, we quantify it as the total annual aggregated energy demand of all hubs in the cluster. Each cluster $m$ receives a benefit of $\frac{\alpha_m}{\sum_m \alpha_m}$ of the total benefit of trading in the network. We assume that $\alpha_m$ is predefined based on the size of the cluster and known for all clusters. Let $P_{m}^{\mathrm{bid}}$ be the vector of energy traded over the complete horizon $\mathcal{T}$ by cluster $m$ and let $C^{\mathrm{bid}}_{m}$ be the net cost bid of the cluster $m$ for the net energy traded over the complete horizon $\mathcal{T}$. 

The resulting weighted bargaining game problem for the clustered network is given by:
\begin{subequations}
\label{eq:bargaining_game}
\begin{align}
    \notag \min_{\substack{\{P^{\mathrm{bid}}_m,  C^{\mathrm{bid}}_m , \Delta J_m\}_{\forall m \in \mathcal{M}} \\ \{p_i, {c}_{i}^{\mathrm{bid}}, \Delta J_i\}_{\forall i \in \mathcal{H}} }} \quad & \sum_{m\in\mathcal{M}} \underbrace{- \alpha_m \ln{(\Delta J_m - C^{\mathrm{bid}}_m)}}_{J^{\mathrm{barg}}_m}\\
    \label{eq:cluster_energy_sum}\text{s.t. } \quad \qquad  & \sum_{m \in \mathcal{M}} P^{\mathrm{bid}}_{m} = 0,\ \  \sum_{m \in \mathcal{M}} C^{\mathrm{bid}}_{m} = 0,\\  
    \label{eq:cluster_hub_sum} & \left \{\begin{array}{l} \begin{aligned}
     p_i \in \mathscr{P}_i, &\ \ \forall i\in \mathcal{H}_m  \\
   \sum_{i\in\mathcal{H}_m} {p}_{i}^{\mathrm{bid}} &= P^{\mathrm{bid}}_{m},\\
   \sum_{i\in\mathcal{H}_m} {c}_{i}^{\mathrm{bid}} &= C^{\mathrm{bid}}_{m},\\
   \sum_{i\in\mathcal{H}_m} {q}_{i}^{\mathrm{bid}} &= 0,\\
    \sum_{i\in\mathcal{H}_m} \Delta J_i &= \Delta  J_{m}\\
    \Delta J_m - & \ C^{\mathrm{bid}}_m \geq 0 \\
\end{aligned} \end{array} \right\} \forall m \in \mathcal{M}
\end{align}
\end{subequations}

The constraint \eqref{eq:cluster_energy_sum} ensures that the electrical energy trades and cost bids of all the clusters sum to 0. This means that no energy or cost is lost during trading with the exception of transmission losses. The explicit computation of network losses as a result of trading and the detailed impact of P2P trading on the network parameters is not considered here. This is left as a topic of future work. Each cluster in the network must also satisfy the local cluster constraint \eqref{eq:cluster_hub_sum}. This ensures that the operational constraints of each hub within the cluster are satisfied. The electrical energy traded by all hubs in a cluster sums to the total cluster energy trade, $P_{m}^{\mathrm{bid}}$. Similarly, the cost bids of hubs in the cluster must add to the total cluster cost bid, $C_{m}^{\mathrm{bid}}$. The thermal energy trades are local to each cluster, hence, the thermal trades of all hubs in a cluster sum to 0. The total cluster cost difference, $\Delta J_m$, is the sum of the individual hub cost differences. The final constraint ensures positivity of $\Delta J_m$ $- C^{\mathrm{bid}}_m$ to enforce the feasibility of the logarithmic objective function and prevent hubs from losing money by participating in the P2P market. In practice, a small positive error term is added within the natural log to avoid numerical issues when solving the optimization problem. The constraint \eqref{eq:cluster_energy_sum} are network constraints, the first constraint in \eqref{eq:cluster_hub_sum} are hub constraints and the rest are cluster constraints for each cluster.

In this study, we propose a nested dual consensus ADMM algorithm \citep{banjac2019decentral} to solve the bargaining game \eqref{eq:bargaining_game} in a distributed, efficient, and scalable manner. The algorithm applies consensus ADMM to the dual of our bargaining game problem. We approach this by reformulating the dual of the problem and solving the following optimization problem for each cluster $m$ in a distributed fashion as discussed below:
\begin{align}
       \notag \min_{\substack{P^{\mathrm{bid}}_m,  C^{\mathrm{bid}}_m, \Delta J_m,\\  \{p_i, {c}_{i}^{\mathrm{bid}}, \Delta J_i \}_{\forall i \in \mathcal{H}_m}}} & - \alpha_m \ln{(\Delta  J_{m} -  C^{\mathrm{bid}}_m)}  + \frac{1}{4\mu M} \lVert \begin{bmatrix}  P^{\mathrm{bid}}_m &   C^{\mathrm{bid}}_m \end{bmatrix}^T + z_{m} \rVert _2^2 \\
       \label{opt:cluster_trading} \text{s.t. } \quad & \eqref{eq:cluster_hub_sum} 
\end{align}
where $z_m$ is a global value for the cluster energy trade and cost bid of cluster $m$ and $\mu_{\geq 0}$ is the step size parameter of the dual consensus ADMM. The objective function considers only the overall cluster cost bid, \(C^{\mathrm{bid}}_{m} = \sum_{i\in\mathcal{H}_m} {c}_{i}^{\mathrm{bid}} \), without specifying how \(C^{\mathrm{bid}}_m\) is distributed among hubs or how \( {c}_{i}^{\mathrm{bid}} \) is computed. For this problem, the optimal dispatch can be determined with any set of hub trading prices \( {c}_{i}^{\mathrm{bid}} \) that sum to \( C^{\mathrm{bid}}_m \), as this choice does not impact the optimal hub and network strategy (proof in \cite{BEHRUNANI20233751}). In this work, we set \( {c}_{i}^{\mathrm{bid}} \) arbitrarily to solve the optimization problem and obtain the optimal cluster and hub setpoints. The specific hub cost bids are then determined by strategically splitting the total cluster bid based on the optimal solution, as summarized in Section \ref{sec:opt}.

The cluster problem \eqref{opt:cluster_trading} is further distributed between the cluster coordinator and the individual hubs and solved using the standard consensus ADMM as presented in Alg. \hyperlink{alg:admm}{3} by creating local copies of the shared variables between each hub and the cluster coordinator. The local variables of the electrical energy trade, thermal energy trade and the cost difference for hub $i$ are denoted by $p^{\mathrm{bid}}_{i,i}$, $p^{\mathrm{bid}}_{i,c}$, and $\Delta J_{i,i}$ and for the cluster coordinator $c$ are denoted by $p^{\mathrm{bid}}_{i,c}$, $q^{\mathrm{bid}}_{i,c}$, and $\Delta J_{i,c}$. The corresponding global variables are $z^{\mathrm{p}}_{i}$, $z^{\mathrm{q}}_{i}$, and $z^{\mathrm{J}}_{i}$, respectively. Additionally, the equality constraints, $p^{\mathrm{bid}}_{i,i} = z^{\mathrm{p}}_{i}$ and $p^{\mathrm{bid}}_{i,c} = z^{\mathrm{p}}_{i}$ are added to the optimization to ensure that the local estimates of the electrical energy trades aligns with global trade value. Similar constraints are also added for the thermal trade and the cost differences. Let $s^{\mathrm{c}}_m$ collect all the shared variables for cluster $m$, $s^{\mathrm{c}}_m = \{ p^{\mathrm{bid}}_{i,c}, q^{\mathrm{bid}}_{i,c}, \Delta J^{\mathrm{bid}}_{i,c} \}_{\forall i \in \mathcal{H}_m}$. Thus, we can write \eqref{opt:cluster_trading} as an optimization problem of the form \eqref{eq:basic_ADMM} that can be solved via Alg. \hyperlink{alg:admm}{3}. The resulting augmented Lagrangian can be separated into the problem of the cluster coordinator and the problem of the individual hubs. 

The cluster coordinator is responsible for ensuring that all the cluster wide constraints are satisfied and for optimizing the objective of the Nash Bargaining Game. The cluster coordinator optimization under an augmented Lagrangian objective is compactly written as:
\begin{align}
        \notag \min_{\substack{ P^{\mathrm{bid}}_m,  C^{\mathrm{bid}}_m\\ \Delta J_m, s^{\mathrm{c}}_m }} & - \alpha_m \ln{(\Delta  J_{m} -  C^{\mathrm{bid}}_m)}  + \frac{1}{4\mu M} \lVert \begin{bmatrix}  P^{\mathrm{bid}}_m &   C^{\mathrm{bid}}_m \end{bmatrix}^T + z_{m} \rVert _2^2 \\[-1.5em]
        \notag &+ \sum_{i\in\mathcal{H}_m} \biggl( \lambda^{\mathrm{p}}_{i,c}   p^{\mathrm{bid}}_{i,c}  + \frac{\rho}{2} \lVert p^{\mathrm{bid}}_{i,c} - z^{\mathrm{p}}_{i} \rVert ^2_2 +\lambda^{\mathrm{q}}_{i,c}   q^{\mathrm{bid}}_{i,c}  + \\
      \label{opt:cluster_trading_coord}   & \quad \quad \quad \frac{\rho}{2} \lVert q^{\mathrm{bid}}_{i,c} - z^{\mathrm{q}}_{i} \rVert ^2_2 + \lambda^{\mathrm{J}}_{i,c}  \Delta J_{i,c}  + \frac{\rho}{2} \lVert \Delta J_{i,c} - z^{\mathrm{J}}_{i} \rVert ^2_2 \biggr)\\
        \notag \text{s.t. } \ \    
        &\sum_{i\in\mathcal{H}_m} {q}_{i,c}^{\mathrm{bid}} = 0,   \sum_{i\in\mathcal{H}_m} {p}_{i,c}^{\mathrm{bid}} = P^{\mathrm{bid}}_{m}, \sum_{i\in\mathcal{H}_m} \Delta J_{i,c} = \Delta  J_{m},
\end{align}
where $\lambda^{\mathrm{p}}_{i,c}$, $\lambda^{\mathrm{q}}_{i,c}$, and $\lambda^{\mathrm{J}}_{i,c}$ are the Lagrange duals variables of the cluster coordinator for the electrical energy trade, thermal energy trade and the cost difference for hub $i$, respectively, and $\rho_{\geq 0}$ is the augmented Lagrangian penalty parameter. 

On the other hand, the hub optimization ensures that the hub constraints are satisfied and the objective function ensures consensus with the cluster coordinator. Let $s_i$ collect all the shared variables for hub $i$, $s_i = \{ p^{\mathrm{bid}}_{i,i}, q^{\mathrm{bid}}_{i,i}, \Delta J^{\mathrm{bid}}_{i,i} \}$. The resulting optimization problem for each hub $i$ in the cluster based on the augmented Lagrangian is given below:
\begin{align}
   \notag     \min_{p_i, s_i}
        &  \quad \lambda^{\mathrm{p}}_{i,i}   p^{\mathrm{bid}}_{i,i}  + \frac{\rho}{2} \lVert p^{\mathrm{bid}}_{i,i} - z^{\mathrm{p}}_{i} \rVert ^2_2 +\lambda^{\mathrm{q}}_{i,i}   q^{\mathrm{bid}}_{i,i}  + \frac{\rho}{2} \lVert q^{\mathrm{bid}}_{i,i} - z^{\mathrm{q}}_{i} \rVert ^2_2\\
  \label{opt:cluster_trading_hub}      & \quad  + \lambda^{\mathrm{J}}_{i,i}  \Delta J_{i,i}  + \frac{\rho}{2} \lVert \Delta J_{i,i} - z^{\mathrm{J}}_{i} \rVert ^2_2 \\
        \text{s.t. } 
   \notag     &\quad  p_i \in \mathscr{P}_i, \ \ \forall i\in \mathcal{H}_m 
\end{align}
where $\lambda^{\mathrm{p}}_{i,i}$, $\lambda^{\mathrm{q}}_{i,i}$, and $\lambda^{\mathrm{J}}_{i,i}$ are the Lagrange dual variables of the hub $i$ for the electrical energy trade, thermal energy trade and the cost difference for hub $i$, respectively.

\begin{figure}[t]
\hypertarget{alg1:dual_descent_bargaining_game}{}
\flushleft
\hrule
\smallskip
\textbf{\textsc{Algorithm 1}}: Dual Consensus ADMM Bargaining game  
\smallskip
\hrule
\smallskip
\textbf{Parameters:} $\alpha_m, \mu, \rho$, Tolerances $\epsilon, \sigma$ \\ [.2em]
\textbf{Initialization:} $\boldsymbol{k}=0$, Cluster duals $y_m^0, d_m^0, \ \forall m \in \mathcal{M}$, \\
\hspace*{5em} $
    \left.\begin{array}{l}
\text{Hub trades }[p^{\mathrm{bid},k}_{i}, q^{\mathrm{bid},k}_{i},  \Delta J_{i}^{k}], \\
\text{Duals }[ \lambda^{\mathrm{p},0}_{i},  \lambda^{\mathrm{q},0}_{i},  \lambda^{\mathrm{J},0}_{i}] \end{array}\right\} \begin{array}{l} \forall i \in \mathcal{H}_m, \\ \forall m \in \mathcal{M} \end{array} $\\
 [.2em]
\textbf{Iterate until convergence:} \scalebox{0.96}{$\left\{\begin{array}{l}\left\|r_m^k\right\|\leq \sigma^{\mathrm{p}},\left\|s_m^k\right\| \leq \sigma^{\mathrm{d}} \ \forall m \in \mathcal{M} \\ \text{\textbf{or} }k\geq k^{\mathrm{max}}\end{array}\right.$} \\[.2em]
$
\left\lfloor
\begin{array}{l}
\text{$\forall m \in \mathcal{M}$ :} \\ \hspace*{0.6em}   
\left\lfloor
\begin{array}{l}
\text{1. Communicate $y_m^k$ with all neighbors $\mathcal{N}_m$.}\\[0.2em]
\text{2. Update dual price:}\\[0.2em]
\qquad d_m^{k+1} \leftarrow d_m^k+\mu \sum_{n \in \mathcal{N}_m}\left(y_m^k-y_n^k\right)
\\ [0.3em]
\text{3. Update cluster trade values:}\\[0.2em]
\qquad 
z_m^{k+1} \leftarrow \mu \sum_{n \in \mathcal{N}_m}\left(y_m^k+y_n^k\right)-d_m^{k+1}
\\ [0.3em]
\text{4. Solve cluster prob. \eqref{opt:cluster_trading_coord} and hub prob. \eqref{opt:cluster_trading_hub} $\forall i \in \mathcal{H}_m$:}\\
\quad \text{using the consensus ADMM Alg. \hyperlink{alg:admm}{3}}\\ [0.2em]
\hspace*{0.8em}\left\lfloor
\begin{array}{l}
\text{\textit{Parameters}: Cluster trade $ z_m^{k+1}$ } \\ 
\text{\textit{Initialize}: Shared variables} \\ \hspace*{4.2em} \text{$[ z^{\mathrm{p},0}_{i},  z^{\mathrm{q},0}_{i}  z^{\mathrm{J},0}_{i}]\leftarrow[p^{\mathrm{bid},k}_{i}, q^{\mathrm{bid},k}_{i},  \Delta J_{i}^{k}$]} \\ \hspace*{4.2em} \text{Duals $[ \lambda^{\mathrm{p},0}_{i},  \lambda^{\mathrm{q},0}_{i},  \lambda^{\mathrm{J},0}_{i}] \leftarrow [ \lambda^{\mathrm{p},k}_{i},  \lambda^{\mathrm{q},k}_{i},  \lambda^{\mathrm{J},k}_{i}] $}\\ 
\text{\textit{Outputs}: Cluster trades and bids $P^{\mathrm{bid},k+1}_m,C^{\mathrm{bid},k+1}_m$ }\\
\hspace*{3.8em} \text{Duals $ \lambda^{\mathrm{p},k+1}_{i}, \lambda^{\mathrm{q},k+1}_{i}, \lambda^{\mathrm{J},k+1}_{i} $}\\
\hspace*{3.8em} \text{Hub setpoints $p_{i}^{k+1}, p^{\mathrm{bid},k+1}_{i}, q^{\mathrm{bid},k+1}_{i},\Delta J_{i}^{k+1}$ }
\end{array}
\right. \vspace*{0.4em}\\ 
\text{5. Compute dual:}\\
\qquad \ y_m^{k+1} \leftarrow \frac{1}{2 \mu M}\left(\left[
P^{\mathrm{bid},k+1}_m \ \  C^{\mathrm{bid},k+1}_m
\right]^T+z_m^{k+1}\right) 
\end{array} \right.\\[1em]
\hspace*{1em}
\vspace*{-0.3em}
\\[-0.6em]
\boldsymbol{k} \leftarrow \boldsymbol{k}+1
\end{array}
\right.$\\[0.3em]
\textbf{Return: } Optimal setpoints $[p_{i}^{\star}]_{\forall i \in \mathcal{H}_m}, P^{\mathrm{bid},\star}_m,C^{\mathrm{bid},\star}_m \forall m \in \mathcal{M}$,\\
\hspace*{3.8em} Duals $y_m^{\star}, d_m^{\star}$ $[ \lambda^{\mathrm{p},\star}_{i},  \lambda^{\mathrm{q},\star}_{i},  \lambda^{\mathrm{J},\star}_{i}]_{\forall i \in \mathcal{H}_m}$ $\forall m \in \mathcal{M}$
\smallskip
\hrule
\end{figure}
The proposed scheme is summarized in Alg. \hyperlink{alg1:dual_descent_bargaining_game}{1} and comprises two nested loops. At iteration $k$, the cluster coordinator communicates the dual of the problem, $y_m^{k}$, with the neighboring clusters in Step. 1, where $\mathcal{N}_m \subseteq \mathcal{M}$ is the set of clusters in the neighborhood of the cluster $m$. It uses the communicated values to locally compute the dual price, $d_m^{k+1}$, in Step 2. and the global cluster trade values, $z_m^{k+1}$, in Step 3.  In the inner loop, consensus ADMM is used within each cluster between the hubs in the cluster and the cluster coordinator in Step 4. to optimize operation of the hubs and compute the energy trades based on the set global cluster trade value. The inner loop outputs a local estimate of the cluster trade and bid values along with the local operational setpoints for the hubs and the duals of the algorithm. To warm start, the inner loop is initialized with its solution from the previous outer loop iteration. The resulting local cluster trade value is then used in Step 5. to compute the new dual, $y_m^{k+1}$, that can be communicated again. This iteration continues until a termination criteria is satisfied. The equilibrium for the optimization is reached and the algorithm is set as converged if the number of iterations of this process, $k$, exceeds a maximum $k^{\mathrm{max}}$ or if the squared norm of primal and dual residuals of each cluster are lower than the tolerances $\sigma^{\mathrm{p}}$ and $\sigma^{\mathrm{d}}$, respectively. Here, the primal and dual residual vectors, $r_{m}^k$ and $ s_{m}^k$ for cluster $m$ respectively, are defined as 
\begin{align}
\notag r_{m}^k & = \frac{1}{2} \left[(y_m^k+y_n^k) \ \forall n \in \mathcal{N}_{m} \right]\\
\notag s_{m}^k & = -\frac{\mu}{2}\left[ (y_m^{k+1}-y_m^k) - (y_n^{k+1}-y_n^k) \ \forall n \in \mathcal{N}_{m} \right].
\end{align}

The results in \cite{banjac2019decentral} prove that the dual descent ADMM algorithm converges. The iteration limit is set here to prevent excessive computation time. If this limit is reached before convergence, we set the inter-cluster trades and cost bids to 0 and restrict hubs to trade only within their cluster until the trades are recomputed. Similarly, consensus ADMM used to solve the intra-cluster trading problem in Step 4 is also guaranteed to converge. If the ADMM algorithm in Step 4 terminates due to the iteration limit before reaching consensus, a disagreement arises among hubs about the trade between them. Each hub trades energy based on its locally computed trade value, leading to mismatched expectations of energy trades. For electrical energy, this mismatch is resolved by purchasing or selling energy to the grid. For thermal energy, it results in thermal comfort violations or additional costs to produce energy within the hub to meet demand, or excess thermal energy being discarded as waste heat.

\section{Distributed MPC and hub cost computation}
\label{sec:opt}

The controller is implemented in a receding horizon fashion. The complete framework for the clustered network optimization and trading price computation is summarized in Alg. \hyperlink{alg:Optimization_pseudocode}{2}. At each time step, the network is reassessed to observe if there are any new clusters that have joined the network or if any hubs have joined or left any of the clusters in Step 2., and the current state of each hub is measured in Step 3. The optimal operation of the clusters is determined at intervals of $t_{\mathrm{rh}}$ times steps in Step 4. Given the network topology, the controller uses Alg. \hyperlink{alg1:dual_descent_bargaining_game}{1} to solve the cluster optimization problem over a prediction horizon of length $T_{\mathrm{cl}}$ to obtain the optimal control input sequence. The cluster trades, $P^{\mathrm{bid},\star}_m$, and bids $C^{\mathrm{bid},\star}_m$, are fixed over the network for the next $t_{\mathrm{rh}}$ time steps. Following this, the optimal cluster trading cost for horizon $T_{cl}$, $C^{\mathrm{bid},\star}_m$, is used to compute an average cluster trading cost for the next $t_{rh}$ times steps, $C^{\mathrm{avg}}_m$, which is what is officially exchanged between clusters. A detailed explanation of computing $C^{\mathrm{avg}}_m$ is in Section \ref{sec:cost_balancing}. This amount is then summed to the total cluster cost $\overline{C}^{\mathrm{bid}}_m$, which is later used to distribute the benefit between hubs within each cluster. Finally, the first input in the computed optimal control sequence is applied to the hubs in Step 5. and the costs are updated in Step 6. This process is repeated every $t_{\mathrm{rh}}$ time steps and the algorithm is executed again to reevaluate the values of $P^{\mathrm{bid}}_{m}$ and $C^{\mathrm{bid}}_{m}$. 
\begin{figure}[!t]
\hypertarget{alg:Optimization_pseudocode}{}
\flushleft
\hrule
\smallskip
\textbf{\textsc{Algorithm 2}}: Distributed MPC optimization  pseudo-code
\smallskip
\hrule
\medskip
\textbf{Initialization: } $t=0$\\ 
\textbf{while} true \textbf{do:}\\ [.3em]
$
\left\lfloor
\begin{array}{l l}
\text{1. If }t \ \% \ t_f = 0 \text{ and } t \neq 0{ :}\\
\hspace*{1.2em}\left\lfloor \begin{array}{l l}
& \hspace*{-1em} \forall m \in \mathcal{M}: \\
 & \hspace*{-1em} \text{i. Compute and settle hub payments, } c^{\mathrm{bid}}_i \ \ \forall i \in \mathcal{H}_m \\
 & \text{for the past $t_f$ hours using $ \overline{C}^{\mathrm{bid}}_m$ in  \eqref{eq:hub_costbalance}}\\ 
 & \hspace*{-1em} \text{ii. Reset } \overline{C}^{\mathrm{bid}}_m  = 0
\end{array}
\right.
\vspace{0.3em}\\
\text{2. Reassess network to update } \mathcal{M}, \text{and} \ \{\mathcal{H}_m\}_{\forall m \in \mathcal{M}}\\
\text{3. Measures the current state $\forall i \in \mathcal{H}$}\\
\text{4. if }t \ \% \ t_{\mathrm{rh}} = 0 { :}\\
\hspace*{1.2em}
\left\lfloor
\begin{array}{l l}
& \hspace*{-1em}\text{i. Compute the $J_{i}^{\mathrm{dec}} \ \ \forall i \in \mathcal{H}$ for the horizon $T_{\mathrm{cl}}$}\\
& \hspace*{-1em} \text{ii. Compute the optimal cluster and hub trades } \\
&  \text{for the complete horizon $T_{\mathrm{cl}}$ using Alg. \hyperlink{alg1:dual_descent_bargaining_game}{1}}\\
& \hspace*{-1em} \text{iii. Fix $P^{\mathrm{bid},\star}_m$ for the next $t_{\mathrm{rh}}$ hours }\forall m \in \mathcal{M}.\\
& \hspace*{-1em} \text{iv. Compute and settle cluster payments,$C^{\mathrm{avg}}_m(t)$, }\\
& \hspace*{0.5em} \text{using \eqref{eq:avg_bid_computation} for the next $t_{\mathrm{rh}}$ hours }\forall m \in \mathcal{M}\\
& \hspace*{-1em} \text{v. Update }\overline{C}^{\mathrm{bid}}_m  \leftarrow \overline{C}^{\mathrm{bid}}_m + C^{\mathrm{avg}}_m(t) \ \forall m \in \mathcal{M}
\end{array} \right. \vspace{0.3em} \\
\hspace*{0.9em}\text{else }t \ \% \ t_{\mathrm{rh}} \neq 0 { :}\\
\hspace*{1.2em}
\left\lfloor
\begin{array}{l l}
& \hspace*{-1em}\text{i. Compute the $J_{i}^{\mathrm{dec}} \ \ \forall i \in \mathcal{H}$ for the horizon $T_{\mathrm{hb}}$}\\
 & \hspace*{-1em} \text{ii. Compute the optimal hub trades/setpoints in each} \\
 & \text{ cluster for horizon $T_{\mathrm{hb}}$ using Alg. \hyperlink{alg1:dual_descent_bargaining_game}{1} for \eqref{opt:cluster_interim_trading} \eqref{opt:hub_interim_trading}}\\ 
\end{array}
\right. \vspace{0.3em} \\ 
\text{5. Apply the optimal control input for time } t \\
\text{6. Update $J_{i}^{\mathrm{grid}}, J_{i}^{\mathrm{dec}} \ \ \forall i \in \mathcal{H}$ for the current $t$}\\
t \leftarrow t+1
\end{array}
\right.
$
\medskip
\hrule
\end{figure}

In Step 4. of \hyperlink{alg:Optimization_pseudocode}{2}, at each time step in the interim, only the inner loop of the Alg. \hyperlink{alg1:dual_descent_bargaining_game}{1}(Step 4.) is executed wherein the hubs within each cluster only recompute their own optimal setpoints with a horizon of length $T_{\mathrm{hb}}$ subject to the previously computed fixed cluster trades. Then, the first time step of the computed control sequence is applied, the response is measured and the process is repeated until $t_{\mathrm{rh}}$ time steps have elapsed and Alg. \hyperlink{alg1:dual_descent_bargaining_game}{1} is executed again. In this case, the cluster coordinator problem no longer needs the bargaining game objective as the trades and bids between clusters are fixed. Therefore, the cluster coordinator problem is solely responsible for the consensus of all the hubs in a cluster wherein its objective contains just the Lagrange dual and ADMM terms. The resulting optimization problem of the coordinator in the inner loop is given below:
\begin{align}
        \notag \min_{s^{\mathrm{c}}_m} & \sum_{i\in\mathcal{H}_m} \biggl( \lambda^{\mathrm{p}}_{i,c}   p^{\mathrm{bid}}_{i,c}  + \frac{\rho}{2} \lVert p^{\mathrm{bid}}_{i,c} - z^{\mathrm{p}}_{i} \rVert ^2_2 +\lambda^{\mathrm{q}}_{i,c}   q^{\mathrm{bid}}_{i,c}  + \frac{\rho}{2} \lVert q^{\mathrm{bid}}_{i,c} - z^{\mathrm{q}}_{i} \rVert ^2_2\\
      \label{opt:cluster_interim_trading}   & \quad \quad \quad + \lambda^{\mathrm{J}}_{i,c}  \Delta J_{i,c}  + \frac{\rho}{2} \lVert \Delta J_{i,c} - z^{\mathrm{J}}_{i} \rVert ^2_2 \biggr)\\
        \notag \text{s.t. } \   
        &\sum_{i\in\mathcal{H}_m} {q}_{i,c}^{\mathrm{bid}} = 0,   \sum_{i\in\mathcal{H}_m} {p}_{i,c}^{\mathrm{bid}} = P^{\mathrm{bid}}_{m}
\end{align}

Similarly, the hub optimization problems \eqref{opt:cluster_trading_hub} also change. In addition to ensuring constraint satisfaction, the optimization also aims to minimize the operational cost of the hubs. The resulting optimization problem of the hub in the inner loop is given below:
\begin{align}
       \notag \min_{p_i, s_i}
        &  \quad J^{\mathrm{grid}}_{i} +\lambda^{\mathrm{p}}_{i,i}   P^{\mathrm{bid}}_{i,i}  + \frac{\rho}{2} \lVert p^{\mathrm{bid}}_{i,i} - z^{\mathrm{p}}_{i} \rVert ^2_2 +\lambda^{\mathrm{q}}_{i,i}   q^{\mathrm{bid}}_{i,i}  + \frac{\rho}{2} \lVert q^{\mathrm{bid}}_{i,i} - z^{\mathrm{q}}_{i} \rVert ^2_2\\
      \notag & \quad  + \lambda^{\mathrm{J}}_{i,i}  \Delta J_{i,i}  + \frac{\rho}{2} \lVert \Delta J_{i,i} - z^{\mathrm{J}}_{i} \rVert ^2_2 \\
      \label{opt:hub_interim_trading}  \text{s.t. } 
        &\quad  p_i \in \mathscr{P}_i, \ \ \forall i\in \mathcal{H}_m 
\end{align}

To ensure that the independent cluster operation is always executable in between cluster bargaining games, the horizon of Alg. \hyperlink{alg1:dual_descent_bargaining_game}{1} has to be long enough so that the cluster trades and bids values always exist for the complete optimization horizon $\mathcal{T}_{\text{hb}}$. Since the cluster optimization is solved every $t_{\mathrm{rh}}$ hours, the following must be fulfilled: 

\begin{equation}
   \notag T_{\mathrm{cl}} \geq t_{\mathrm{rh}} + T_{\mathrm{hb}}
\end{equation}

\begin{figure}[t]
\centering
\includegraphics[width=89mm]{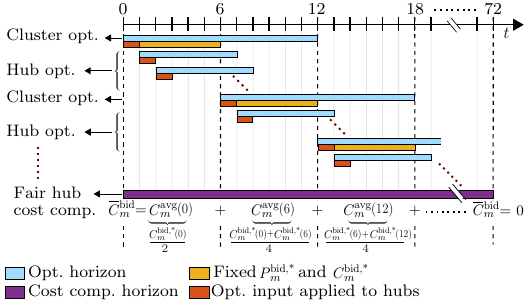}
  \caption{An example of this complete optimization framework executed over an interval of 72h with $T_{\mathrm{cl}} = 12$h and $t_{\mathrm{rh}} = T_{\mathrm{hb}} = 6$h and a sampling time of 1h. The computation of the average $C^{\mathrm{avg},\star}_m$ fo every time interval is shown and the hub cost distribution is done at the end with $t_{\mathrm{f}} = 72$h.}
  \label{fig:alg_timing}
 \end{figure}

Both $T_{\mathrm{cl}}$ and $T_{\mathrm{hb}}$ are integral multiples of $t_{\mathrm{rh}}$, i.e., $T_{\mathrm{cl}} = \zeta_{\mathrm{cl}} t_{\mathrm{rh}}$ where $\zeta_{\mathrm{cl}} \in \mathbb{Z}_{\geq 1}$. To speed up convergence at each time step, the optimal trade and dual values from the previous time step are used to initialize the algorithms. The receding horizon repetition brings feedback into the process through the measurements and allows the controller to continuously adapt to new forecast information, suppress the effect of model mismatch and disturbances, as well as anticipate increasing or decreasing energy prices. Fig. \ref{fig:alg_timing} shows an example of this control scheme applied over an interval of 72h with $T_{\mathrm{cl}} = 12$h and $t_{\mathrm{rh}} = T_{\mathrm{hb}} = 6$h with a sampling time of 1h. The time scale separation between the cluster and hub optimization enables the optimization of energy trades with fair compensation in the complete network while minimizing the total computation since the cluster coordination happens only at set time intervals and remains fixed in between. Meanwhile, the hubs can still locally adapt at each time step and participate in energy trades with hubs in other clusters without coordinating with other clusters. 

\subsection{Average cluster bid and hub cost computation}\label{sec:cost_balancing}

The optimal trading cost of the barganing game for cluster $m$, $C^{\mathrm{bid},\star}_m$, computed for the complete $T_{\mathrm{cl}}$, cannot be used directly since the electricity trades between clusters are fixed for only the first $t_{\mathrm{rh}} < T_{\mathrm{cl}}$ time steps before being recomputed. The proportion of the $C^{\mathrm{bid},\star}_m$ that is fixed for $t_{\mathrm{rh}}$ before it is recomputed is equal to $C^{\mathrm{bid},\star}_m \cdot (t_{\mathrm{rh}}/T_{\mathrm{cl}}) = (C^{\mathrm{bid},\star}_m/\zeta_{\mathrm{cl}})$. However, if we simply chose to use $C^{\mathrm{bid},\star}_m/\zeta_{\mathrm{cl}}$ as the trading value for each $t_{\mathrm{rh}}$, later changes in the energy trading trajectories, and thus changes in the cost incurred by each cluster, would be ignored. Instead, an average cluster trading cost, $C^{\mathrm{avg}}_m$, is computed every $t_{\mathrm{rh}}$ time steps each time the cluster trade optimization is solved, which is paid by the clusters for the trading that occurs over the next $t_{\mathrm{rh}}$ time steps. The average cost is computed by averaging the portion of the trading cost from the current bargaining game that influences \( t_{\mathrm{rh}} \), given by \( C^{\mathrm{bid},\star}_m(t)/\zeta_{\mathrm{cl}} \), along with the trading costs from previous bargaining games over the same time interval:  
\[
\frac{C^{\mathrm{bid},\star}_m(t - t_{\mathrm{rh}})}{\zeta_{\mathrm{cl}}}, \ \ \frac{C^{\mathrm{bid},\star}_m(t - 2 t_{\mathrm{rh}})}{\zeta_{\mathrm{cl}}}, \ \  \dots, \ \  \frac{C^{\mathrm{bid},\star}_m(t - (\zeta_{\mathrm{cl}} - 1) t_{\mathrm{rh}})}{\zeta_{\mathrm{cl}}}
\]
This approach incorporates both recent and past bargaining game outcomes within the given interval, ensuring that delayed cluster benefits from the unfixed portion of the horizon are accounted for in each game. Fig. \ref{fig:alg_timing} shows an example of how the average $C^{\mathrm{avg}}_m(t)$ is computed every $t_{\mathrm{rh}}$ hours by averaging all the the previously computed $C^{\mathrm{bid},\star}_m(t)$ values for that interval. This average is expressed explicitly using the following equation:

%This average cost combines the computed trading costs from the fixed portion of the current bargaining game, $C^{\mathrm{bid},\star}_m(t)/\zeta_{\mathrm{cl}}$, and the unfixed portions of previous bargaining games over the same time interval. For each time interval, $\zeta_{\mathrm{cl}}$ values of past $C^{\mathrm{bid},\star}_m(t)$ exist and are used to compute the average. 

%\begin{equation}
%    \label{eq:avg_bid_computation}C^{\mathrm{avg}}_m(t) = \left \{\begin{array}{l l}  \begin{aligned}
%        \sum_{s = 0}^{(t-t_{\mathrm{rh}})/t_{\mathrm{rh}}}\frac{C^{\mathrm{bid},\star}_{m}(s . t_{\mathrm{rh}})}{\zeta_{\mathrm{cl}} \cdot (t/t_{\mathrm{rh}})} , \ \ \ & \text{if } t \leq T_{\mathrm{cl}} \\
%         \sum_{s = (t-T_{\mathrm{cl}})/t_{\mathrm{rh}}}^{(t-t_{\mathrm{rh}})/t_{\mathrm{rh}}}\frac{C^{\mathrm{bid},\star}_{m}(s . t_{\mathrm{rh}})}{\zeta_{\mathrm{cl}}^2} \ \ \ & \text{otherwise.}
%         \end{aligned}
%       \end{array} 
%       \right.
%\end{equation}

\begin{equation}
    \label{eq:avg_bid_computation}C^{\mathrm{avg}}_m(t) = 
         \frac{1}{\zeta_{\mathrm{cl}}} \sum_{s = 0}^{\zeta_{\mathrm{cl}}-1} \frac{ C^{\mathrm{bid},\star}_{m}(t - s . t_{\mathrm{rh}}) }{\zeta_{\mathrm{cl}}}
\end{equation}
To compute the average at the start of the algorithm, when $t \leq T_{\mathrm{cl}}$, only the available values are used:  
\begin{equation}
\notag C^{\mathrm{avg}}_m(t) = 
         \frac{1}{(t/t_{\mathrm{rh}})} \sum_{s = 0}^{t/t_{\mathrm{rh}} - 1}\frac{C^{\mathrm{bid},\star}_{m}(t - s  . t_{\mathrm{rh}})}{\zeta_{\mathrm{cl}}}.
\end{equation}

The average cluster costs that are settled between clusters are further distributed between the hubs of each cluster. This hub cost distribution need not occur each time the average cost is computed, but may be accumulated over a long time horizon and settled all at once. This allows the cost payments between hubs of a cluster to occur less frequently rendering time scale flexibility and reducing computation when unnecessary. In this study, we consider that the cost balancing between hubs takes place every $t_{\mathrm{f}}$ time steps for the trades that occurred in the past $t_{\mathrm{f}}$ time steps, where $t_{\mathrm{f}}$ is an integer multiple of $t_{\mathrm{rh}}$. The cost distribution is done using $\overline{C}^{\mathrm{bid}}_m$ which sums $C^{\mathrm{avg}}_m(t)$ calculated every $t_{\mathrm{rh}}$ time steps over the last $t_{\mathrm{f}}$ time steps. Once the hub cost distribution is complete, $\overline{C}^{\mathrm{bid}}_m$ is reset to $0$ to record it from scratch for the next $t_f$ hours. In the example presented in Fig. \ref{fig:alg_timing}, $t_{\mathrm{f}} = 72$h is used.  The cluster cost is distributed by the cluster coordinator between the hubs of the cluster to ensure that all hubs have the same relative cost benefit compared to the decentralized cost when no trading occurs, that is, all hubs benefit equitably by participating in the market. This is achieved by solving the following optimization:
\begin{align}
\notag \min_{\beta_m, \{c^{\mathrm{bid}}_i\}_{\forall i \in \mathcal{H}_m}} & \beta_m \\
\label{eq:hub_costbalance}\text { s.t. } \quad & \sum_{i \in \mathcal{H}_m} c^{\mathrm{bid}}_i= \overline{C}^{\mathrm{bid}}_m,\\
\notag & \beta_m  = \frac{\big(J_{i}^{\mathrm{grid}}+c^{\mathrm{bid}}_i\big)-J_{i}^{\mathrm{dec}}}{J_{i}^{\mathrm{dec}}}, \forall i \in \mathcal{H}_m 
\end{align}
where $\beta_m$ is the relative cost benefit achieved by each hub in the cluster. The optimization determines $\beta_m$ and the optimal trading cost, $c^{\mathrm{bid}}_i$, for each hub $i$ in the cluster. The constraints ensure that relative cost benefit is equal throughout the cluster and the sum of all the hub trading cost is equal to the total $\overline{C}^{\mathrm{bid}}_m$ paid by the cluster coordinator over the last $t_{\mathrm{f}}$ time steps.

\section{Plug-and-play strategies}\label{sec:pnp}

Network growth and topology changes are the main driver for developing P2P energy markets that support plug-and-play. The network should be able to support a hub joining or leaving a cluster. Similarly, it should also be possible for an entire cluster to join or leave the network. It is crucial these situations does not have an impact on the entire network. In other words, the other hubs in the network are able to optimize and operate without disruption or additional unfair costs. Here, we consider the plug-and-play procedures for the following different scenarios. 

\subsection{Cluster Plug-in and Plug-out}

In this study, clusters are higher-level entities that comprise multiple hubs. Therefore, we assume that a cluster does not go online or offline unexpectedly, and that cluster-level changes are always scheduled ahead of time. We also assume that all hubs will not plug-out of a cluster simultaneously, which is essentially a cluster plug-out. Outage situations that may unexpectedly impact one or more clusters are not considered here. Cluster changes are scheduled to occur only right before cluster trading occurs, every $t_{\mathrm{rh}}$ time steps. The set of clusters in the network remains fixed otherwise. If a cluster plans to join the network and trade energy with the hubs in other clusters in the network, they are operated independently until the bargaining game is solved again. When $t\%t_{\mathrm{rh}} = 0$ again, the cluster is added to the set $\mathcal{M}$. Similarly, if a cluster plans to leave the network and no longer participate in P2P trading with the hubs in the other clusters, the previously computed cluster trades $P^\mathrm{bid}_m$ and the cluster payment $C^\mathrm{avg}_m (t)$ must be fulfilled until the next cluster problem is solved next, that is, until $t\%t_{\mathrm{rh}} = 0$ again. 

\subsection{Hub Plug-in and Plug-out}

If a hub joins or leaves a cluster, the change can be directly incorporated into the inner-loop optimization  of the corresponding cluster without impacting other clusters. The cluster coordinator controller (\eqref{opt:cluster_trading_coord} or \eqref{opt:cluster_interim_trading}) is reconfigured to adjust electrical and thermal trade accordingly. Additionally, the joining or leaving hub has its own optimization problem (\eqref{opt:cluster_trading_hub} or \eqref{opt:hub_interim_trading}) which The hub is then added to or removed from the distributed optimization for future time steps.
Since each hub is connected to the electricity grid, the hubs are still always able to fulfill their electricity demand and the fixed cluster electricity trade. Similarly, in case of a hub exiting the cluster, any potential thermal trade with that hub is compensated internally through conversion or storage in the absence of trading. This ensures that the equality constraints can always be satisfied, and the cluster coordinator and the hub optimization problems do not become infeasible due to the departure of a single hub. Prior to the next instance of the cluster bargaining game being solved(when $t \% t_{\mathrm{rh}} = 0$), the weights of the bargaining game, $\alpha_m$, are also reevaluated and communicated to the clusters by the network operator to adapt to the updated network structure. Since the cluster bid is not recomputed until then, this may have an impact on the net benefit of each hub in the cluster. This impact, however, is limited to $t_\mathrm{rh}$-1 time steps. A smaller value for $t_\mathrm{rh}$ reduces this impact as the hub is accounted in the inter-cluster bargaining game solution sooner. This is illustrated by the simulations presented in Section~\ref{sec:PnP_impact_sim}.

\subsubsection{Impact on intra-cluster cost distribution}

When a new hub joins the cluster, its decentralised cost \( J_{i}^{\mathrm{dec}} \) and grid cost \( J_{i}^{\mathrm{grid}} \) are calculated only for the duration of its participation in the cluster. At the end of \( t_\mathrm{f} \) time steps, the total cluster trading cost, \( \overline{C}^{\mathrm{bid}}_m \), is distributed among hubs using \eqref{eq:hub_costbalance}, and excludes the costs of hubs that were not part of the cluster during that period. This approach ensures that hubs can integrate seamlessly into the cluster and are compensated fairly based solely on the time that they are actually involved in the trading network.

In contrast, the impact of the hub leaving has to be included in the cluster cost distribution to ensure that it does not negatively impact or completely diminish the benefit of the other hubs in the cluster and thereby, de-incentivize them remaining in the cluster. Therefore, the cluster cost distribution problem presented in \eqref{eq:hub_costbalance} is modified. The decentralized cost of the plugged out hub $i$, $J_{i}^{\mathrm{dec}}$ is split into when the hub was still plugged into the cluster, $J_{i}^{\mathrm{dec,in}}$, and after it is plugged out of the cluster, $J_{i}^{\mathrm{dec,out}}$. $J_{i}^{\mathrm{grid}}$ is only computed as part of the optimization when the hub was still plugged into cluster and hence, is denoted here by $J_{i}^{\mathrm{grid,in}}$. To ensure that the remaining hubs in the cluster profit from the arrangement, the hub $i$ that leaves might be charged a potential penalty, $c^{\mathrm{pen}}_{i}$, proportional to the decentralized cost from when they left the cluster. This makes sure that when multiple hubs leave a cluster, the hubs are penalized in accordance to their relative size, contribution and the total time that they were not part of the cluster. The optimization objective also comprises a regularization term for the total penalty $\gamma$. The penalty is $0$ for the hubs that stay within the cluster since $J_{i}^{\mathrm{dec,out}}$ is $0$ for these hubs. The resulting modified cost distribution optimization is given below:
\begin{align}
\notag \min_{\substack{\beta_m, \gamma,  \\ \{ c^{\mathrm{bid}}_i, c^{\mathrm{pen}}_i\}_{\forall j \in \mathcal{H}_m}}} & \beta_m + 
 W \left\|\gamma\right\|_2^2\\
\notag \text { s.t. } \quad & \sum_{i \in \mathcal{H}_m} c^{\mathrm{bid}}_i + \gamma = \overline{C}^{\mathrm{bid}}_m,\\
\notag & \beta_m \leq \beta^{\mathrm{max}}\\
& \beta_m = \frac{\big(J_{i}^{\mathrm{grid}}+c^{\mathrm{bid}}_i\big)-J_{i}^{\mathrm{dec,in}}}{J_{i}^{\mathrm{dec,in}}},  \\
\notag & c^{\mathrm{pen}}_i = \frac{J^{\mathrm{dec,out}}_{i}}{\sum_{k \in \mathcal{H}_m}J^{\mathrm{dec,out}}_{k}} \cdot \gamma, \ \  \forall i \in \mathcal{H}_m,
\end{align}
 
The above constraints ensure that in the case of hub plug-out, the hubs can achieve a relative cost reduction lower than $\beta^{\mathrm{max}}$. In this study, we set $\beta^{\mathrm{max}} = 0$ to ensure that the cost achieved with trading is not higher than cost without trading. This value can also be set lower to ensure a minimum cost reduction for the hubs. The hub trading costs and the total penalty are set accordingly. The total cluster cost, $\overline{C}^{\mathrm{bid}}_m$, is constituted not only of the trading costs of all the hubs but also of the total penalty $\gamma$ levied to the hubs leaving the cluster. Finally, the last constraint distributes the penalty between the different hubs plugging out proportional to the total $J^{\mathrm{dec,out}}$ of the cluster. 

\section{Numerical Simulations}
\label{sec:experiments}

In this section, the proposed method is simulated on various networks of diverse energy hubs to analyze the resulting behavior and computational scalability.  Decentralized MPC method, using optimization formulation \eqref{eq:dec_energyhub_optimization} is used as a benchmark for each simulated case. The performance of the proposed controller is also compared to the centralized MPC (CMPC) presented in \eqref{eq:energyhub_optimization} as well as a na\"ive distributed MPC (DMPC) control strategy. A baseline $9$-hub network divided into $3$ clusters of $3$ hubs each is used to analyze the trading behavior, seasonal results, and PnP scenarios. Further simulations use networks of up to 18 hubs grouped into different numbers of clusters to demonstrate the scalability of the proposed method. Each energy hub has distinct electrical and thermal demand trajectories built with real annual building data from ETH Z{\"u}rich and EMPA campuses. The hubs configurations are based on either the real capacities and configurations present therein or by designing new hubs with device capacities sized according to peak demand and to encourage trading. Details of the technologies present within each hub and the corresponding capacity and constraint limits for all the technologies are available on Gitlab \cite{koepele2024code}. We assume that energy hubs have access to perfect weather and demand forecasts. Ambient temperature and solar irradiation data are provided by ETH Z{\"u}rich and EMPA weather stations.

Each energy hub is connected to the public electricity and gas grids and receives the planned prices from the DSO. The electricity prices vary between on-peak and off-peak values whereas the gas price and electricity feed-in tariff are constant over time. Table \ref{tab:pnp_tariff} specifies the tariffs for utilizing grids, which is based on Swiss prices. Hubs in a cluster are additionally connected by a local thermal grid. All numerical simulations were performed in python using the solver MOSEK to solve bargaining game optimization problems, and the solver GUROBI to computes the solution of the remaining optimization problems. All simulations are performed on the ETH Z{\"u}rich Euler cluster. %Figure \ref{fig:elect_grid_hub} illustrates the available pathways for each energy resource in an energy hub cluster. 

 \begin{table}[!h]
    \centering
    \begin{tabular}{l c c}
                \hline
                Tariff & Parameter & Cost(CHF/kW) \\ \hline
                Electricity - peak/offpeak & {$c_{\text{in,e}}$} & 0.27/0.22\\
                Electricity - feed-in  &$c_{\text{out,e}}$& 0.12\\ 
                Gas &$c_{\text{g}}$ & 0.115\\
                Electricity grid &$c_{\text{tr}}$& 0.02\\
        \end{tabular}
        \caption{Prices for electricity and gas usage.}
                \label{tab:pnp_tariff}  
\end{table}

The simulation periods vary between $3$ and $8$ days in length and the sampling time is $1$~h. The prediction horizon of the cluster optimization, $T_{\mathrm{cl}} = 24$h with $t_{\mathrm{rh}} = 12$h. Hence, $\zeta_{\mathrm{cl}} = 2$. Similarly, the prediction horizon of the intra-cluster hub optimization, $T_{\mathrm{hb}} = 12$h. For each simulation, the hub costs are computed only at the end of the complete simulation period of either $3$ or $8$ days, hence $t_{\mathrm{f}} = 72$h or $t_{\mathrm{f}} = 198$h, respectively. For the primal consensus ADMM algorithm, tolerances $\epsilon^p$ and $\epsilon^d$ are set to 0.05 and 0.03 respectively. The maximum number of iterations allowed, $w^{\mathrm{max}}$, is set to 200. These values are selected based on the work in \cite{behrunani2023distributed}. The tolerances for the dual consensus ADMM to solve the bargaining game, $\sigma^{\mathrm{p}}$ and $\sigma^{\mathrm{d}}$, are both set to 0.003, and the maximum number of iterations allowed for the algorithm, $k^{\mathrm{max}}$, is set to 200. The step sizes for primal and dual consensus ADMM vary to encourage convergence in fewer iterations. The primal consensus ADMM step size, $\rho$, is initialized at a value of 0.001 and increases at a rate of $2\%$ after every iteration. The dual consensus ADMM step size, $\mu$, is initialized at 2000 and decreases at a rate of $3\%$ after every iteration. In our study, the maximum iterations is never reached and the algorithm always converges by being below the tolerance. 

\subsection{Modified Day Ahead Bargaining Game} 

The cost and convergence results from a single day-ahead bi-level bargaining solution from the proposed method are presented in this section. The cost values reported below represent the actual costs that would be incurred by the hubs during real-time operation by applying the optimal control input at each time step. These include the planned costs of purchasing gas and electricity from the grid, as well as trading costs. The results of the weighted bargaining game proposed here are compared to the standard unweighted game for a 3-hub network divided into $3$ clusters. The three hubs vary in annual demand where Hub $1$ has the most and Hub $3$ has the least. This annual demand  of each cluster is used to determine weights $\alpha_1 = 123$, $\alpha_2 = 26$, and $\alpha_3 = 4$ for the weighted bargaining game. In practice, these values would be provided by the DSO. The weights are set to $\alpha_1 = \alpha_2 = \alpha_3 = 1$ in the unweighted case. %The isolated bargaining game cost is compared to that of the day-ahead Decentralized solution \textcolor{red}{(Perhaps reference the optimization problems solved from the methods section)}. 

\begin{figure}[!t]
     \centering
     \includegraphics[width=\linewidth]{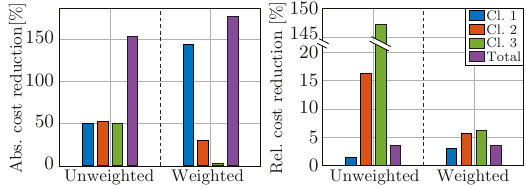}
     \caption{Comparison of the absolute and relative cost reduction achieved by the complete network and each cluster in the network using the standard and the weighted bargaining game formulations.}
     \label{fig:bargaining_fairness}
 \end{figure}

Fig. \ref{fig:bargaining_fairness} presents the resulting absolute cost benefit, $J_{m} - J_{m}^{\mathrm{dec}}$, and relative cost benefit, $100\% \times (J_{m} - J_{m}^{\mathrm{dec}})/J_{m}^{\mathrm{dec}}$, of each single-hub cluster $m$ compared to the decentralized solution cost when no P2P trading occurs in the network. The unweighted bargaining game distributes the total benefit equally between each cluster which in turn results in a highly disproportionate relative cost reduction. Such a strategy may also discourage some clusters to participate in the trading scheme as their cost benefit is relatively negligible to their overall cost, such as for Cluster 1 in this example. Conversely, the costs in the weighted bargaining game are redistributed according to the assigned weights that pulls the relative benefit towards equality. This ensures that clusters of varying sizes benefit equitably proportional to their size and contribution and matches the overall social cost reduction as much as possible. Since the cluster weights are computed based on the annual demand, the selected weights do not yield the exact same relative benefits in a simulated 24-hour period out of the whole year. The relative cluster benefits are expected to trend towards equality over longer simulations. 

\begin{figure}[!t]
     \centering
     \includegraphics[width=\linewidth]{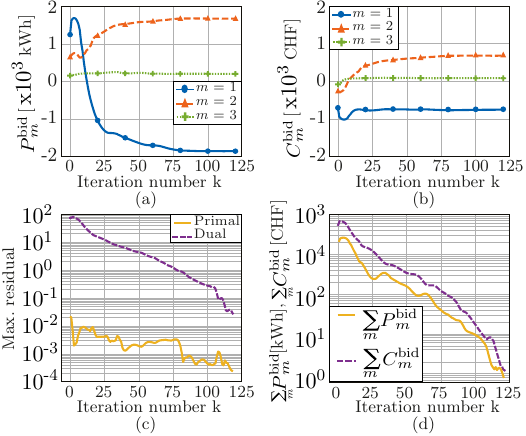}
     \caption{Convergence results of the dual ascent ADMM algorithm used to solve the bargaining game. Evolution of (a) cluster energy trade $P_{m}^{\mathrm{bid}}$, (b) cluster cost bid $C_{m}^{\mathrm{bid}}$, (c) maximal residuals and (d) sum of cluster energy trades, $\sum_m P_{m}^{\mathrm{bid}}$ and the cluster costs, $\sum_m C_{m}^{\mathrm{bid}}$ as the number of iterations increase.}
     \label{fig:bargaining_convergence}
 \end{figure}

Figure \ref{fig:bargaining_convergence} shows the convergence of the clusters' bi-level bargaining game. Fig. \ref{fig:bargaining_convergence}(a) and (b) show how the cluster trades and cluster trading cost bids settle to their final value within 75 iterations after which the values refine slightly in the further iterations until the primal and dual residuals (shown in Fig. \ref{fig:bargaining_convergence}(c)) is within the tolerance limit. Finally, 
Fig. \ref{fig:bargaining_convergence}(d) shows how the sums of all trades and cluster costs evolve over time and converges to zero (amounts to $<1$ kWh and $<1$ CHF). 

% Normal Operation: Trading, optimality, comp time & iterations, seasonality
\subsection{Case Study on a 9-Hub Clustered Energy Hub Network}

\begin{figure}[!t]
     \centering
     \includegraphics[width=\linewidth]{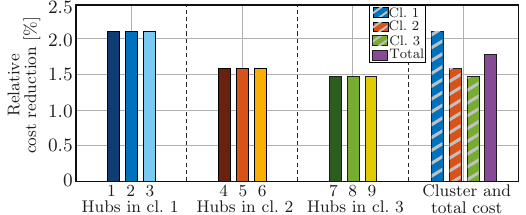}
     \caption{Optimal cost reduction for each hub and each cluster in the network computed using the proposed algorithm.}
     \label{fig:bargaining_fairness_hubs_all}
 \end{figure}

This section will discuss the results from simulating the 9 hub network in Fig. \ref{fig:network_cluster} with the proposed method for $3$ days. Cluster 1 comprises Hubs 1, 2, and 3, Cluster 2 comprises Hubs 4, 5, and 6, and Cluster 3 comprises the remaining 3 hubs. Fig. \ref{fig:bargaining_fairness_hubs_all} illustrates the relative cost reduction achieved by each cluster in the network and the hubs within each cluster compared to the decentralized cost when no trading occurs. Each cluster receives a similar benefit due to the weighted bargaining game. Furthermore, the cluster benefits are distributed fairly amongst the hubs in the cluster and as a result, every hub in the cluster receives the same benefit as the whole cluster that matches the overall cluster cost reduction. The accumulated bargaining costs for each cluster are divided between its energy hubs as described in Section \ref{sec:cost_balancing} at the end of the simulation. Overall, the network obtains a benefit of $1.78\%$ over the simulation period and cluster benefits range from 1.48\% for cluster 3 to a benefit of 2.1\% achieved by cluster 1.

\begin{figure}[!t]
     \centering
     \includegraphics[width=\linewidth]{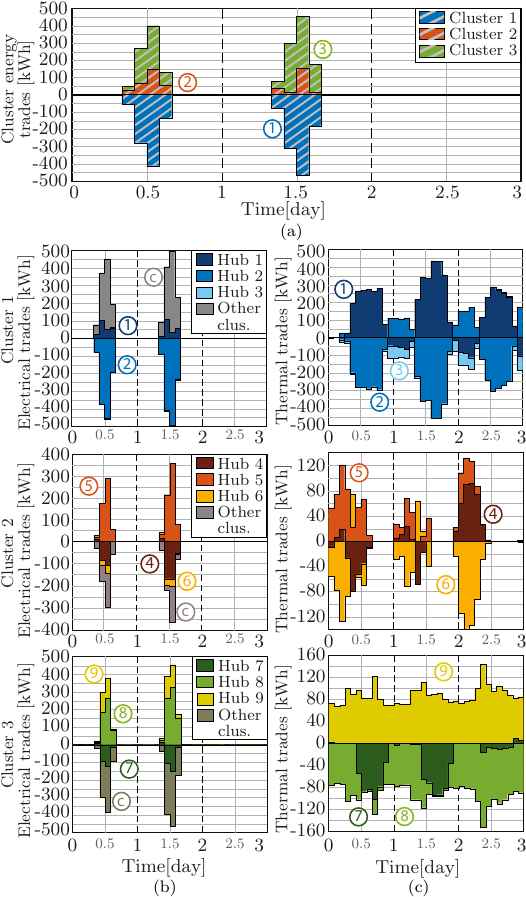}
     \caption{Total (a) Electrical energy trades between the three clusters over a period of 3 days and the total (b) Electrical and (c) Thermal energy traded by the hubs in each cluster.}
     \label{fig:trades_all}
 \end{figure}

The electricity and heat trading contributions between hubs and clusters are shown in Fig. \ref{fig:trades_all}. A negative trade indicates energy is exported by the hub/cluster and vice versa. We observed that electricity trading does not occur on the last day due to a lack of solar irradiance. Overall, it is shown in Fig. \ref{fig:trades_all}(a) that Cluster $1$ predominantly exports electricity to the other clusters. This is particularly due to Hub $2$, which has the highest electricity production capacity and most of the energy produced is transferred to external clusters (as seen in Fig. \ref{fig:trades_all}(b)). Cluster $2$ only takes a small amount of electricity from the other clusters which indicates it is internally well-balanced with its demands. Cluster $3$ has a much lower electricity generation capacity, particularly Hub $8$, and therefore utilizes the cluster network to optimally fulfill its demand. Fig. \ref{fig:trades_all}(b) shows that most of the electrical energy imported into Cluster 3 from other clusters is consumed by Hub 8. The hubs can also trade thermal energy internally within each cluster to fulfill their thermal demand as seen in Fig. \ref{fig:trades_all}(c). 

The resulting cost bids are shown in Table \ref{tab:trading_costs}. These costs cover both electrical and thermal energy being traded within each cluster as well as the electricity traded between clusters. The hubs that provide energy are paid by those that consume the traded energy. The bargaining game implicitly modifies the total cost to cover the varying cost of energy at different times and attain a fair relative benefit across all clusters.

 \begin{table}[!t]
        \centering
    \begin{tabular}{l r|l r|l r}
        \hline
         \multicolumn{6} {c} {Trading Costs (CHF)}\\ \hline
         Hub 1&$283$& Hub 4&$-205$&Hub 7&$-187$\\
         Hub 2&$-816$&Hub 5&$312$&Hub 8&$171$\\
         Hub 3&$-42$&Hub 6&$-68$&Hub 9&$552$\\\hline
         \textbf{Cluster 1}&$\bm{-575}$&\textbf{Cluster 2}&$\bm{39}$&\textbf{Cluster 3}&$\bm{536}$\\
    \end{tabular}
    \caption{Total trading bids for each hub and cluster over a period of 3 days.}
                \label{tab:trading_costs}  
                
\end{table}

% PnP Results
\subsection{PnP Scenarios and comparison to other control methods}
\label{sec:PnP_impact_sim}
This section will analyze how the proposed solution performs on the baseline network under different scenarios of hubs plugging into and out of the network. The simulation is $3$ days long, with the PnP event occurring at $6$ AM on the second day. For each scenario, a hub is removed from or added to the original network. Hubs operate independently using the decentralized controller when they are not participating in the network. We examine the scenarios when a producing hub, Hub $2$, is added or removed, and when a consuming hub, Hub $8$, is added or removed from the configuration. 

\begin{figure}[t]
     \centering
     \includegraphics[width=\linewidth]{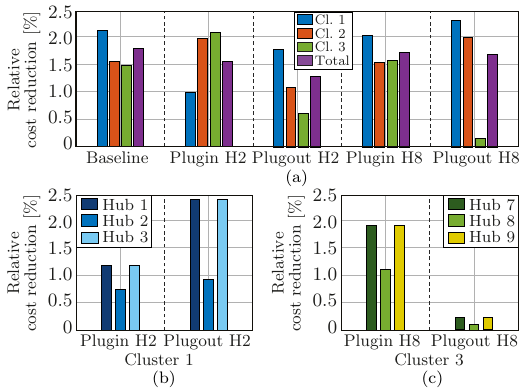}
     \caption{Simulation results for (a) the complete network under different PnP scenarios and the impact on the hubs in (b) Cluster 1 and (c) Cluster 3 when a specific hub in the cluster is plugged into/out of the network. }
     \label{fig:pnp_plug_in}
 \end{figure}

Fig. \ref{fig:pnp_plug_in} shows the relative cost reduction for hubs and clusters over the whole simulation time for the PnP scenarios and the baseline case without PnP. For each hub PnP scenario, all clusters and all hubs receive a net-benefit for participating in the network compared to if they were islanded. Additionally, the hub that is plugging in or out of the network obtains less of a relative benefit than the other hubs in its cluster. This is shown in Fig. \ref{fig:pnp_plug_in} (b) and (c) for Hub 2 and Hub 8, respectively.  This is expected, since the mobile hub spends less time than its neighbors in the market. We observe that plugging in a hub yields a lower overall network cost, and therefore results in a higher cost reduction, compared to plugging the same hub out. This is because the trading flexibility within a cluster is expanded rather than reduced to fulfill the required cluster trading.   

We now analyze the behavior of each PnP case in more detail. Cluster 2 and Cluster 3 perform better than the baseline case when a producing Hub 2 from Cluster 1 is plugged in. In this case, less weight was given to Cluster 1 during the bargaining game since it did not anticipate Hub 2 joining. Hubs 1 and 3 obtain less benefit than the baseline case as a result shown in \ref{fig:bargaining_fairness_hubs_all}. The joining hub 2
On the other hand, the producing hub exiting the network actually benefits its neighboring hubs in the cluster. The weights in the preceding inter-cluster bargaining game reserve an absolute benefit considering Hub 2, which is divided among fewer hubs later after Hub 2 leaves. The remaining hubs are able to adjust their energy strategy to exceed the baseline benefit and produce more energy or purchase more electricity form the grid to fulfill the cluster trade commitment. The producing Hub 2 leaving also impacts other clusters in the next iteration of the inter-cluster bargaining game once the weights are readjusted. It results in a lower benefit for Cluster 2 and 3 since the energy that was previously imported by the clusters from Cluster 1 now has to be sourced at a higher price from the electricity grid. In each of these two cases, Hubs 1 and 3 benefit more than Hub 2 since the cluster benefits are no longer split equally among the hubs and are distributed in accordance to the relative time spent in the cluster. 

When the consuming hub is plugged in, neighboring hubs actually fare better than the baseline scenario shown in Fig. \ref{fig:bargaining_fairness_hubs_all}. This is because Hub 8 produces and trades thermal energy to the hubs in the cluster and increases the overall energy flexibility of the cluster despite the cluster weight being outdated. In contrast, all hub benefits in Cluster 3 drop significantly when Hub 8, a consuming hub, exits the network. This is because as a consuming hub, Hub 8, was paying for the electricity and once it is plugged out, the remaining hubs in Cluster 3 are left with a cluster trading fee and excess electricity, which they sell back to the grid. Similar to Cluster 1, Hubs 7 and 9 always benefit more than Hub 8 as it is not present in the cluster the for the complete simulation. Overall, the hubs are able to plug in and plug out of the network easily and the cost distribution of the clusters ensures that the hubs are compensated fairly based on the time within the cluster. The hubs are then also incorporated in the next instance of the bargaining game and the inter-cluster benefits adjust accordingly. 

\begin{figure}[t]
     \centering
     \includegraphics[width=\linewidth]{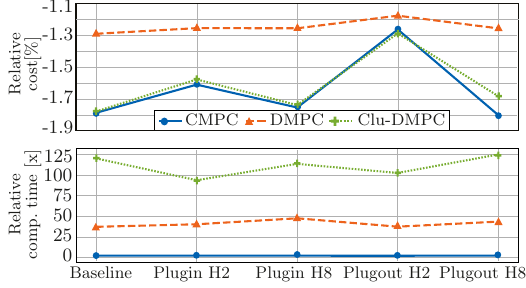}
     \caption{Comparison of the proposed controller with CMPC and standard DMPC under different PnP scenarios in terms of relative cost (\%) and relative computation time ($\times$), with respect to the decentralized solution without trading.}
     \label{fig:cost_compare}
 \end{figure}

Finally, Fig. \ref{fig:cost_compare} displays the relative cost and computation time ratio with respect to the decentralized control scheme of the proposed controller as well as of the centralized controller and a na\"ive distributed MPC (DMPC) wherein Alg. \hyperlink{alg:admm}{3} is used directly to distribute the CMPC optimization problem without any network clustering or cost bids. The termination conditions for DMPC are the same as those used for the primal consensus ADMM in our strategy, with a constant step size $\rho = 0.04$. To ensure a fair cost comparison, thermal trading in CMPC and DMPC is restricted to neighboring hubs that are within the same cluster during the clustered setup. The total network costs of the proposed method are close to optimal CMPC for the baseline case and all PnP scenarios. The cost achieved by the clustering scheme is lower than the cost using DMPC, which has a higher cost sub-optimality with the trade-off that the DMPC has a lower computation time. Adjusting the tolerance of the DMPC can further reduce the cost and bring it closer to the optimal CMPC cost. However, this will negatively impact the computation time of DMPC, as it will require more time to reach consensus. The results would be similar to the proposed method and take away the computation time advantage of the DMPC. The added benefit of using the proposed controller is that the clusters operate independently from one another for a majority of the time, thereby maintaining more privacy, and do not need to communicate with all other clusters in the network to compute the optimal energy trades within the cluster. The computation time of all the controllers remains relatively similar across scenarios for each controller. The proposed controller (Clu-DMPC) requires more time than the CMPC. This is due to the presence of nested iterative loops in the distributed bargaining game algorithm that require the small hub problem to be solved multiple times to achieve the optimal operation in contrast to the CMPC where a single large optimization problem is solved just once. However, we expect that the computation time for the proposed algorithm will approach that of CMPC as the network size increases. Furthermore, from the scalability study in \cite{behrunani2023distributed}, we would expect that the computation times of the proposed distributed method would be more competitive to that of centralized methods for more accurate - but challenging - energy hub optimization formulations, like MILP. We use simplified energy hub models in this study which results in a linear program optimization problems that can be solved relatively easily for small networks. 

\subsection{Seasonal behavior and Scalability}

\begin{figure}[t]
     \centering
     \includegraphics[width=\linewidth]{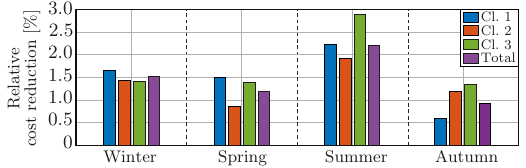}
     \caption{Simulation results for the network over an 8-day period under varied weather conditions.}
     \label{fig:seasons}
 \end{figure}

To analyze the benefits and performance of the proposed algorithm in various circumstances, the baseline network is simulated for a period of $8$ days in each season. The relative cost benefits for each cluster and the entire network in each season are shown in Figure \ref{fig:seasons}. Each cluster—and by extension, each hub—experiences a relative cost reduction similar to its neighbors across seasons. algorithm consistently benefiting the network and remaining robust to external conditions. The results show that the algorithm consistently benefits the network and is impervious to external conditions. The network benefits most from trading in the extremes of winter and summer weather. This is due to higher thermal demands in the winter and more solar irradiation in the summer, which results in an increase in trading in these periods. In each season, the relative cost reduction of each cluster in the network is close to the total network cost reduction that shows that the benefits are distributed equitably among the clusters.

\begin{figure}[t]
     \centering
     \includegraphics[width=\linewidth]{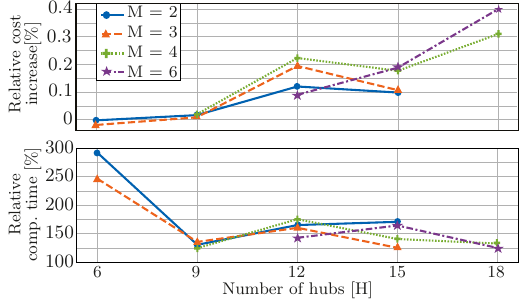}
     \caption{Scalability of the proposed method for networks with different numbers of hubs and clusters in terms of relative cost (\%) and relative computation time ($\times$) with respect to the CMPC.}
     \label{fig:scalability}
 \end{figure}

We analyze the scalability of the proposed controller by simulating the system with different network sizes and topologies for a period of $3$ days. Fig. \ref{fig:scalability} shows the relative cost increase and computation time ratio of the proposed method with respect to the CMPC solution as the number of hubs in the network increases from 6 to 18 and the hubs are divided into 2, 3, 4 or 6 clusters. The hubs are divided into clusters to have at least 2 hubs in each cluster to have nearly the same number of hubs in each cluster. The cost of each network is close to the CMPC solution and sub-optimality increases marginally as the network size increases. With more hubs in a network, there is a larger number of trades that are determined in a distributed fashion which could contribute to the relative cost increase. The computation time of the proposed method is significantly greater than CMPC for small networks as also seen before in Fig. \ref{fig:cost_compare} due to the presence of nested iterative loops in the distributed controller. However, the computation time for the proposed algorithm is expected to approach that of CMPC as the network size increases. Indeed, Fig. \ref{fig:scalability} shows that the computation time ratio with respect to CMPC is generally decreasing and remains nearly equal for different cluster topologies for networks of more than 9 hubs. This is because the solution time grows exponentially for single optimization problems with more decision variables. In contrast, distributed methods solve hub optimization problems in parallel, so the number of hubs in a network impacts the computation time only slightly. However, it is not strictly decreasing since the hubs are not uniform and which hubs are clustered together impacts the convergence for both the periodic bargaining algorithm and the per-hour distributed algorithm. 

\section{Conclusion} \label{sec:conclusion}

This work presents a clustering-based P2P energy trading framework for the optimal operation and energy trading in energy hub networks while enabling plug-and-play (PnP) functionality. By structuring energy hubs into clusters, the proposed method localizes controller updates, reducing computational burden and ensuring scalability. A bi-level game was formulated, where inter-cluster energy trading and cost commitments were determined at the higher cluster level through a Nash Bargaining Game solved using dual consensus ADMM, while energy dispatch and cost-sharing between hubs were optimized at the lower intra-cluster level. This separation allows clusters to operate independently with distributed MPC when cluster trading commitments are fixed, and limits impact to the affected cluster coordinator during PnP events. The hub-level optimization is also solved in a distributed manner using ADMM, ensuring computational feasibility while preserving privacy. Simulations on a diverse network of hubs demonstrate that the proposed algorithm achieves network costs close to the optimal centralized MPC solution, with all hubs benefiting similarly from trading. Additionally, the PnP procedures effectively handle network topology changes, allowing hubs to join or leave with minimal disruption while maintaining overall benefits. While the proposed algorithm requires more time to solve than centralized methods due to its nested and distributed iterations, it is expected to be more advantageous for larger networks and when more complex energy hub models are used. Future work aims to demonstrate this advantage on larger networks with more sophisticated energy hub models using longer simulation times, and experimentally validate the results presented here. We aim to incorporate grid and building models and network losses in the problem and analyze the impact of clustering strategies, and inexact demand and price forecasts on the performance.

\section*{Acknowledgement}
This research is supported by the SNSF through NCCR Automation (Grant Number 180545) and the Swiss Federal Office of Energy (SFOE).  

\bibliographystyle{elsarticle-num-names}
\bibliography{references.bib}
\appendix
\newpage
\section{Distributed optimization - Consensus ADMM}
\label{Appendix:A}

Consider the general global consensus optimization problem comprising of $N$ agents each with local variable $x_i \in \mathbb{R}^n$ that should be equal, decoupled convex objective function $f_i(x_i)$, and  $x_i$ must lie within a local constraint set $\mathcal{X}_i$. This problem can be rewritten with a common global variable $z$: 
\begin{align}
\notag \min_{x_1, \dots,x_N } & \sum_{i=1}^N f_i\left(x_i\right) \\
\label{eq:basic_ADMM} \text { s.t. } & x_i-z=0, \\
 \notag & x_i \in \mathcal{X}_i, \quad i=1, \ldots, N 
\end{align}
The global variable $z$ ensures that the local $x_i$ align. ADMM for the problem can be derived from the augmented Lagrangian given below:
\begin{equation}
\notag L_\rho\left(x_{1, \ldots, N}, z, \lambda_{1, \ldots, N}\right)=\sum_{i=1}^N\left(f_i\left(x_i\right)+\lambda_i^T\left(x_i-z\right)+\frac{\rho}{2}\left\|x_i-z\right\|_2^2\right)
\end{equation}
The augmented lagrangian renders \eqref{eq:basic_ADMM} seperable. The resulting Consensus ADMM algorithm is presented in Alg. \hyperlink{alg:admm}{3} ~\citep{boyd:ADMM}. The algorithm solves the local problem and the solution is communicated to the other agents to update the global estimate of the coupling variables $z$ and dual variables.  This alternation continues until the local
and global values converge. The primal and dual residuals for consensus ADMM in each iteration $w$, $r^w$ and $s^w$, respectively, are defined as:
\begin{equation}
\notag r^w = [x_i^w-z^w \ \forall i \in\mathbb{Z}_{[1,N]}]^{T} , \quad s^w = \rho [z^w-z^{w-1} \ \forall i \in\mathbb{Z}_{[1,N]}]^{T}
\end{equation}
Convergence is achieved when the squared norms of primal and dual residuals are less than the tolerances $\epsilon^p$ and $\epsilon^d$, respectively, or if the iteration count $w$ exceeds the maximum iterations $w^{\mathrm{max}}$.
\begin{figure}[t] 
\hypertarget{alg:admm}{}
\flushleft
\hrule
\smallskip
\textbf{\textsc{Algorithm 3:} } Consensus ADMM
\smallskip
\hrule
\smallskip
\textbf{Initialization: } $\boldsymbol{w}=0$, $ z^0,  \lambda^{0}_i \  \forall i \in \mathbb{Z}_{[1,N]} $ \\ [.2em]
\textbf{Iterate until convergence:}  $w \geq w^{\text{max}}$\textbf{ or } $\left\|r^{w}\right\|^{2}_{2} \leq \epsilon^p, \left\|s^{w}\right\|^{2}_{2} \leq \epsilon^d  $
\\[.2em]
$
\left\lfloor
\begin{array}{l}
\forall i \in \mathbb{Z}_{[1,N]}:\\
\hspace*{.5em} \left\lfloor
\begin{array}{l}
\text{1. Solve: }\scalebox{0.90}{$x_i^{w+1} =\operatorname{argmin}_{x_i}\left(f_i\left(x_i\right)+\lambda_i^{Tw}\left(x_i-z^w\right)+ \frac{\rho}{2}\left\|x_i-z^w\right\|_2^2\right)$}\\ [.2em]
\text{2. Communicate $x_i^{w+1}$ to all other agents.}\\  [0.2em]
\text{3. Update primal: } z^{w+1} =\frac{1}{N} \sum_{i=1}^N x_i^{w+1} \\[0.2em]
\text{4. Update dual: } \lambda_i^{w+1} =\lambda_i^k+\rho\left(x_i^{w+1}-z^{w+1}\right) \\
\end{array} \right.\\[1em]
\hspace*{1em}
\vspace*{-0.3em}
\\[-0.6em]
w \leftarrow w+1
\end{array}
\right.$\\[0.3em]
\textbf{Return: } $x_i^{\star} \leftarrow z^{w}, \lambda_i^{\star}\leftarrow \lambda_i^{w+1} \ \forall i \in \mathbb{Z}_{[1,N]}$
\medskip
\hrule
\end{figure}

\end{document}